\documentclass[aps, prb, twocolumn,amsmath,amssymb,floatfix,reprint,longbibliography, nofootinbib]{revtex4-2}
\usepackage{graphicx}
\usepackage{physics}
\usepackage{times}
\usepackage{dcolumn}
\usepackage{hyperref} 
\hypersetup{colorlinks, allcolors=blue}
\allowdisplaybreaks
\begin{document}

\title{Phonon-mediated intrinsic topological superconductivity in Fermi arcs}

\author{Kristian M{\ae}land}
\email[Contact author: ]{kristian.maeland@uni-wuerzburg.de}
\author{Masoud Bahari}
\author{Bj{\"o}rn Trauzettel}
\affiliation{Institute for Theoretical Physics and Astrophysics, University of W{\"u}rzburg, D-97074 W{\"u}rzburg, Germany}
\affiliation{Würzburg-Dresden Cluster of Excellence ct.qmat, D-97074 W{\"u}rzburg, Germany}

\begin{abstract}
We propose that phonons can intrinsically mediate topological superconductivity on the surface of Weyl semimetals. Weyl semimetals are gapless topological materials with nondegenerate zero energy surface states known as Fermi arcs. We derive the phonon spectrum and electron-phonon coupling in an effective model of a Weyl semimetal and apply weak-coupling Bardeen-Cooper-Schrieffer theory of superconductivity. In a slab geometry, we find that surface superconductivity dominates over bulk superconductivity in a range of chemical potentials around the Weyl nodes.  The superconducting gap function realizes spinless chiral $p$-wave Cooper pairing in the Fermi arcs, leading to Majorana bound states in the core of vortices. Furthermore, we find a suppression of the absolute value of the gap in the center of the Fermi arcs, which is not captured by a local Hubbard attraction. The suppression is due to the nonlocal origin of electron-phonon coupling, leading to a layer dependence which has important consequences for topological surface states. 
\end{abstract}

\maketitle

\section{Introduction}
Topological materials hold great promise for future technological applications due to their protection from small perturbations. Weyl semimetals are gapless topological materials with zero-energy surface states 
called Fermi arcs. The Fermi arcs connect the projection of Weyl nodes with opposite chirality \cite{Yan2017WeylRev, Armitage2018WeylRev}. 
Recent experiments suggest superconductivity occurs in the Fermi arcs of  PtBi$_2$, MoTe$_2$, and TaIrTe$_4$, representing essentially two-dimensional (2D) surface superconductivity \cite{Kuibarov2024FermiArcSCNat, Schimmel2023ExpWeylSCSTM, Hoffmann2024PtBi2STM, Veyrat2023ExpWeyl2DSC, Huang2025PtBi2STM, Moreno2025PtBi2vortex, Naidyuk2018WSC, Xing2020WSC}.

Several theoretical \cite{Cho2012WSC, Wei2014WSC, Hosur2014WSC, Bednik2015WSC, Alidoust2017WSC} and experimental \cite{Qi2016WSC, vanDelft2020Apr, Shipunov2020ExpWeylSC, Bashlakov2022PtBi2phononExp, Zabala2024WeylBulkSC} studies consider superconductivity in the bulk of doped 3D Weyl semimetals. 
More recently, Ref.~\cite{Nomani2023FermiArcSCDOS} showed that there will always be a region of chemical potentials where surface superconductivity dominates due to a much larger surface density of states (DOS) compared to the bulk \cite{Bai2025WSMSC}.

In Ref.~\cite{Trama2024TRSWeylSM_SC}, the authors assume a local Hubbard attraction and predict an anisotropic $s$-wave spin-singlet gap with maximum amplitude in the center of the Fermi arc.
The structure of the gap function has also been discussed from symmetry arguments \cite{Vocaturo2024PtBi2Effective, Waje2025PtBi2GL}. While a local Hubbard attraction is often a good model for phonon-mediated pairing, the underlying mechanism of superconductivity in Weyl semimetals remains largely unexplored. In this paper, we study phonon-mediated pairing in detail and show that it has a distinct form to a local Hubbard interaction.
The electronic states of Fermi arcs are most surface localized in the center of the arc and become gradually more bulk like as they approach the projection on the bulk Weyl node. 
We show that since electron-phonon coupling arises from a Taylor expansion of the hopping term, the interaction is not completely local, and the degree to which electron states penetrate into the bulk affect the coupling strength. Therefore, phonons can give a local minimum in the absolute value of the gap function in the center of the arc, and maxima instead placed between the center of the Fermi arc and its endpoints.

A theoretical study of the underlying mechanism of superconductivity in Fermi arcs can shed light on the possible nontrivial topology of the superconducting state \cite{Kuibarov2024FermiArcSCNat, Naidyuk2018WSC, Xing2020WSC}.
Topological superconductors are highly sought after for use in topological quantum computing \cite{TopoQuantumCompRevModPhys, Leijnse2012TSCrev, Bernevig2013, TopoSCrevSato}.
Topological superconductivity was first discussed for a system of spinless fermions \cite{Kitaev2001Oct}. Since electrons have spin, physical realizations are challenging \cite{Frolov2021Apr}. Fu and Kane \cite{Fu2008Kane} showed that proximity induced $s$-wave superconductivity on the surface of a topological insulator (TI) results in a spinless chiral $p$-wave superconductor in the band basis, thanks to the spin-orbit coupling (SOC) in the TI and the nondegenerate nature of the surface band. The difference from usual spinless chiral $p$-wave topological superconductors is the retained time-reversal symmetry (TRS) \cite{Fu2008Kane}. However, from such a superconducting state, topologically protected Majorana bound states may be engineered in the core of vortices after breaking TRS with a small out-of-plane magnetic field, realizing 2D topological superconductivity \cite{Fu2008Kane, Fukui2010FuKaneTopoStability}. 
Such extrinsic methods of realizing topological superconductivity, involving a topologically trivial superconductor brought in contact with a nonsuperconducting material often having a strong SOC, remain the most common \cite{Oreg2010Oct, SkTopoSCNagaosa, TopoSCrevSato, TopoSCandSkRev, Frolov2021Apr}.

Meanwhile, intrinsic mechanisms of topological superconductivity, where the pairing mechanism directly favors topologically nontrivial gap functions, are rare. Such a realization is attractive since there is a potential for higher operating temperatures and larger gap amplitudes. Various materials with broken inversion symmetry or strong SOC or both have been suggested as candidates of intrinsic topological superconductivity \cite{TopoSCrevSato, Brydon2014CuBiSe, Zhang2021intrinsicTSC, vonRohr2023intrinsicTSC, Kim2023YPtBi, Bahari2024Jun}. 
Several studies discuss the role of electron-phonon coupling in topological superconductivity \cite{Brydon2014CuBiSe, Scheurer2016NCSTSC, Li2023intrinsicTSC} and other topological phase transitions \cite{Cangemi2019PhTopo, Islam2024PhTopo, Bhattacharyya2024PhTopo, Lahiri2025PhTopo}.
Furthermore, magnons or spin fluctuations can generate intrinsic topological superconductivity in heterostructures \cite{Maeland2023AprPRL, Maeland2023DecTSC, VinasBostrom2024May, SunMaeland2023Aug, Sun2024May, Thingstad2025Feb, Lundemo2024May}. 

In this paper, we show that phonons mediate chiral $p$-wave superconductivity in the nondegenerate Fermi arc surface states of Weyl semimetals, representing an intrinsic realization of the Fu and Kane model \cite{Fu2008Kane}. 
In Sec.~\ref{sec:Electrons}, we describe the effective hexagonal crystal model of Weyl semimetals introduced in Ref.~\cite{Vocaturo2024PtBi2Effective}.
With open boundary conditions (OBC) in the $z$ direction, Fermi arcs on top and bottom surface appear in the first Brillouin zone (1BZ). The model is designed to have Fermi arcs in similar positions as PtBi$_2$ and display the same symmetries \cite{Vocaturo2024PtBi2Effective}. 
We then derive the phonon spectrum in Sec.~\ref{sec:phonon}, the electron-phonon coupling in Sec.~\ref{sec:EPC}, and the phonon-mediated electron-electron interaction in Sec.~\ref{sec:SC}, all within the hexagonal crystal model. Working in the band basis, the effective electron-electron interaction has a chiral $p$-wave momentum dependence originating with SOC. As shown in Sec.~\ref{sec:SC}, a generalization of Bardeen-Cooper-Schrieffer (BCS) theory of superconductivity \cite{BCSshort, BCS} then predicts a gap function with a chiral $p$-wave form. We conclude in Sec.~\ref{sec:conclusion}, and the Appendixes provide further details of the model and the competition of superconductivity between bulk and surfaces.


\section{Electrons} \label{sec:Electrons}
Trigonal PtBi$_2$ crystallizes in the P31m space group, has broken inversion symmetry, and is time-reversal symmetric \cite{Shipunov2020ExpWeylSC, Kuibarov2024FermiArcSCNat, Vocaturo2024PtBi2Effective}. There are three unique layers, each of which are triangular lattices with three atomic bases. Following Ref.~\cite{Vocaturo2024PtBi2Effective}, we consider an effective model of PtBi$_2$ on a hexagonal crystal with a single atom in the basis and two orbitals per atom. The electron Hamiltonian is
\begin{equation}
    H_{\text{el}} = -\sum_{i\ell\sigma} (\mu+\mu_\ell)c_{i\ell\sigma}^\dagger c_{i\ell\sigma} + H_{\text{hop}} + H_{\text{SOC}} + H_{\gamma}. 
\end{equation}
The chemical potential $\mu$ controls the doping, $\mu_\ell$ is an orbital-dependent onsite energy, and $c_{i\ell\sigma}^{(\dagger)}$ destroys (creates) an electron at site $i$ in orbital $\ell$ with spin $\sigma$. We do not specify the orbitals, and name them $\ell = A,B$. We set $\mu_A = -\mu_o, \mu_B = \mu_o$, the lattice constant $a=1$, and Planck's constant $\hbar = 1$ throughout. The hopping term is
$
    H_{\text{hop}} = -\sum_{i\Bar{\boldsymbol{\delta}} \ell \ell' \sigma} t_{\ell\ell'}(\Bar{\boldsymbol{\delta}}) c_{i+\delta,\ell\sigma}^\dagger c_{i\ell'\sigma},
$
with $\Bar{\boldsymbol{\delta}} \in \{ \Bar{\boldsymbol{\delta}}_i\}_{i=1}^{8}$ representing the nearest-neighbor vectors. 
For in-plane hopping we use $t$ for intraorbital hopping and $t_o$ for interorbital hopping. For out-of-plane hopping, $\beta$ parametrizes the strength; see Appendix \ref{app:Weylmodel} for details.

We partially Fourier transform these terms to momentum space, with periodic boundary conditions (PBC) in the $x$ and $y$ directions and OBC in the $z$ direction (a slab geometry). This is achieved by letting 
$
    c_{i\ell\sigma} = (1/\sqrt{N_L})\sum_{\boldsymbol{k}} c_{\boldsymbol{k} z_i \ell\sigma}e^{i\boldsymbol{k}\cdot \boldsymbol{r}_i},
$
where $N_L$ is the number of sites per layer and $z_i = 1,2,\dots,L$ labels the $L$ layers from bottom to top. Throughout this paper we use a bar over 3D vectors, like $\Bar{\boldsymbol{r}}_i = (x_i,y_i,z_i)$ for the position of lattice site $i$, while we use no bar over 2D vectors, like $\boldsymbol{r}_i = (x_i, y_i)$.
The SOC term is 
\begin{align}
    H_{\text{SOC}} &= \sum_{\boldsymbol{k}z_i\ell} \big[ s_{\boldsymbol{k}} \big( c_{\boldsymbol{k} z_i \ell \uparrow}^\dagger c_{\boldsymbol{k} z_i \bar{\ell}\downarrow} - \frac{1}{2}\sum_{\delta = \pm1}  c_{\boldsymbol{k} z_i \ell\uparrow}^\dagger c_{\boldsymbol{k}, z_i+\delta, \bar{\ell}\downarrow} \big)\nonumber\\
    & +\text{H.c.}\big],
\end{align}
where $s_{\boldsymbol{k}} = \alpha [ \sin k_y + \cos(\sqrt{3}k_x/2)\sin(k_y/2) -i \sqrt{3} \sin(\sqrt{3}k_x/2)\cos(k_y/2)]$, $\bar{\ell} = B(A)$ when $\ell = A(B)$, and H.c.\ denotes the Hermitian conjugate of the preceding term. The SOC factor $s_{\boldsymbol{k}}$ has a chiral $p$-wave momentum dependence in the 1BZ of the 2D triangular lattice, typical of Rashba SOC. 
Finally, $H_{\gamma} = \sum_{\boldsymbol{k} z_i \ell \sigma} \gamma c_{\boldsymbol{k} z_i \ell\sigma}^\dagger c_{\boldsymbol{k} z_i  \bar{\ell}\sigma}$ breaks inversion symmetry \cite{Vocaturo2024PtBi2Effective}.

\begin{figure}
    \centering
    \includegraphics[width=0.99\linewidth]{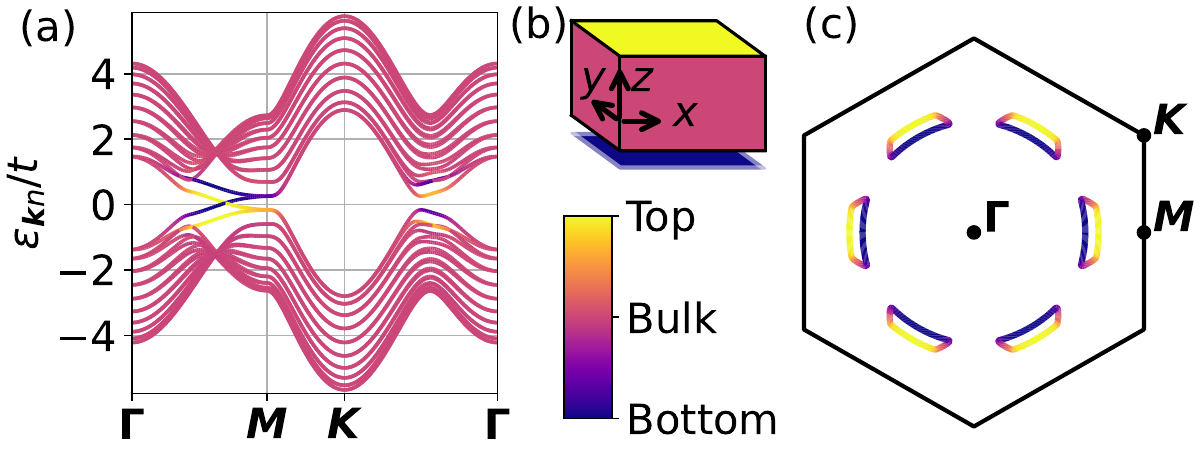}
    \caption{(a) Electron bands in a slab geometry, with each band colored by the weight of its eigenstate $W_{\boldsymbol{k}n}$. Note two nondegenerate surface states crossing the Fermi level between $\boldsymbol{\Gamma}$ and $\boldsymbol{M}$. (b) Illustration of the system and the color code for top surface, bulk, and bottom surface. (c) Fermi surface (FS) shown within the 1BZ, colored by the weight of the eigenstate. The full FS contains the Fermi arcs on the two surfaces and the projection of the bulk FS. The parameters are $t_o/t = 1.5$, $\beta/t = -1.5$,  $\mu/t = -0.05, \mu_o/t = 0.2, \alpha/t = -0.18$, $\gamma/t = -0.2$, (a) $L=10$, and (c) $L = 40$.  }
    \label{fig:OBCweight}
\end{figure}

As detailed in Appendix \ref{app:Weylmodel}, the electron Hamiltonian is gathered in a $4L \times 4L$ matrix which we diagonalize to find the $4L$ bands $\epsilon_{\boldsymbol{k}n}$. The bands are shown in Fig.~\ref{fig:OBCweight}(a) colored by the weight of their eigenstates, defined as
$
    W_{\boldsymbol{k}n} = \sum_{z_i = 1}^{L} (z_i-1)|\psi_{\boldsymbol{k}n,z_i}|^2/(L-1),
$
where $|\psi_{\boldsymbol{k}n,z_i}|^2$ is the sum of squares of the four entries in the eigenvector of band $n$ associated with layer $z_i$. PtBi$_2$ has its Weyl nodes placed about $50$~meV above the Fermi level \cite{Kuibarov2024FermiArcSCNat}, so we mostly consider $\mu/t = -0.05$ in this paper, assuming $t = 1$~eV. Figure~\ref{fig:OBCweight}(c) shows the Fermi surface (FS) at that doping, revealing separate Fermi arcs on the surfaces and projections of the bulk FS.

\section{Phonons} \label{sec:phonon}
We derive the phonon spectrum in the hexagonal crystal through a force constant approach assuming small ionic displacements from equilibrium \cite{BruusFlensberg, Klogetvedt2023, Syljuasen2024, Thingstad2020Jun, Leraand2025Feb}. By applying the symmetries of the P31m space group and considering up to next-nearest-neighbor potentials, we limit the description to three free parameters $\gamma_1, \gamma_3,$ and $\gamma_6$, see details in Appendix \ref{app:phonon}. We treat them as phenomenological parameters and choose values to get a phonon spectrum with a reasonable range of energies \cite{Lucas1968PhononOBC, Benedek2010PhononOBC, Bashlakov2022PtBi2phononExp}.  
Working with a single atomic basis we get three acoustic phonons with 3D PBC, while in the slab geometry there are three acoustic and $3L-3$ optical phonons, in the sense that they have a nonzero energy at zero momentum \cite{Lucas1968PhononOBC, Benedek2010PhononOBC}. 
It is worth noting that PtBi$_2$, with a nine-atomic basis, has three acoustic and 24 optical phonon branches with 3D PBC. So, within the effective model we are catching the three acoustic modes, which given their lower energy should be expected to make the main contribution to superconductivity.

\begin{figure}
    \centering
    \includegraphics[width=0.95\linewidth]{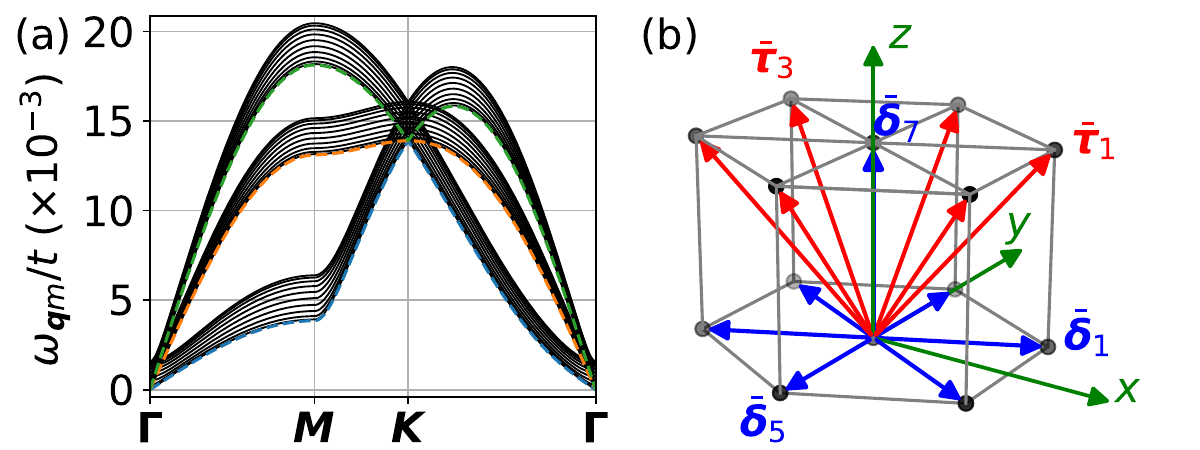}
    \caption{(a) Black lines show the phonon spectrum in a slab geometry. Blue, orange and green dashed lines show the three acoustic phonon modes with periodic boundary conditions (PBC) in all directions at $q_z = 0$. (b) Hexagonal crystal with illustrations of some nearest ($\bar{\boldsymbol{\delta}}_i$) and next-nearest-neighbor vectors ($\bar{\boldsymbol{\tau}}_i$). The parameters are $\gamma_1 = -(0.005t)^2$, $\gamma_3 = 0.45\gamma_1$, $\gamma_6 = 1.5 \gamma_1$, and $L = 10$.}
    \label{fig:phonon}
\end{figure}

The force constant approach involves constructing a dynamical matrix in momentum space which is $3L\times 3L$ in the slab geometry. The phonon spectrum $\omega_{\boldsymbol{q}m}$ is the square roots of the eigenvalues of the dynamical matrix while its eigenvectors $\hat{e}_{\boldsymbol{q}m}$ describe the phonon modes. 
The phonon Hamiltonian is then
$
    H_{\text{ph}} = \sum_{\boldsymbol{q}m} \omega_{\boldsymbol{q}m} a_{\boldsymbol{q}m}^\dagger a_{\boldsymbol{q}m},
$
where $a_{\boldsymbol{q}m}^{(\dagger)}$ annihilates (creates) a phonon in mode $m$ with momentum $\boldsymbol{q}$.
Experimental investigations of the phonon spectrum in PtBi$_2$ indicate a peak in the phonon DOS around $15$~meV \cite{Bashlakov2022PtBi2phononExp}. If $t= 1$~eV, then the phonon spectrum in Fig.~\ref{fig:phonon}(a) should give a large DOS in a similar energy range.

\section{Electron-phonon coupling} \label{sec:EPC}
We derive the electron-phonon coupling (EPC) by Taylor expanding the hopping terms around small ionic displacements \cite{Thingstad2020Jun, Leraand2025Feb}. The ionic displacements are subsequently quantized in terms of the phonon operators, and a Fourier transform yields
\begin{equation}
\label{eq:HEPC}
    H_{\text{EPC}} = \sum_{\boldsymbol{k}\boldsymbol{q}m} g_{\boldsymbol{k}+\boldsymbol{q}, \boldsymbol{k}}^{m} (a_{-\boldsymbol{q},m}^\dagger+a_{\boldsymbol{q}m})d_{\boldsymbol{k}+\boldsymbol{q}}^\dagger d_{\boldsymbol{k}}.
\end{equation}
We have dropped the band index on the electron band operators $d_{\boldsymbol{k}}$, focusing only on the band that has an FS, which is the one relevant for superconductivity. Here,
$g_{\boldsymbol{k} \boldsymbol{k}'}^{m} = \sum_{\ell\ell'\sigma z_i \delta_z} g_{\boldsymbol{k} \boldsymbol{k}' m}^{\ell\ell' z_i \delta_z}v_{\boldsymbol{k},z_i+\delta_z,\ell,\sigma}^* v_{\boldsymbol{k}' z_i \ell'\sigma},$
where $c_{\boldsymbol{k} z_i \ell\sigma} = v_{\boldsymbol{k} z_i \ell\sigma}d_{\boldsymbol{k}}$ and
\begin{align}
\label{eq:gEPC}
    g_{\boldsymbol{k} \boldsymbol{k}' m}^{\ell\ell' z_i \delta_z} &= \sum_{\boldsymbol{\delta}} \frac{\chi t_{\ell\ell'}(\Bar{\boldsymbol{\delta}})}{\sqrt{2N_L M\omega_{\boldsymbol{k}-\boldsymbol{k}',m}}}(e^{-i\boldsymbol{k}'\cdot \boldsymbol{\delta}}\Bar{\boldsymbol{e}}_{\boldsymbol{k}-\boldsymbol{k}', m}^{z_i+\delta_z}\nonumber\\
    &-e^{-i\boldsymbol{k}\cdot\boldsymbol{\delta}}\Bar{\boldsymbol{e}}_{\boldsymbol{k}-\boldsymbol{k}',m}^{z_i})\cdot \Bar{\boldsymbol{\delta}}.
\end{align}
Here, $M$ is the mass of the ions which in this system of units is an inverse energy $M = Ma^2/\hbar^2$. 
To set a value for $M$, we use $M = 204$~u and $a = 1$~\AA. 
$\Bar{\boldsymbol{e}}_{\boldsymbol{q} m}^{z_i}$ is the vector of length 3 associated with layer $z_i$ from $\hat{e}_{\boldsymbol{q}m}$. 
Note that the behavior of the factor in square brackets is different for $\delta_z = 0$ and $\delta_z = \pm 1$, 
i.e., whether the origin is in-plane or out-of-plane hopping terms.
Hence, we discuss in-plane and out-of-plane type EPC as two separate contributions.
$\chi$ is a dimensionless number inversely proportional to the standard deviation of the atomic orbitals \cite{Thingstad2020Jun, Leraand2025Feb}. Appendix \ref{app:EPC} provides more details of the derivation of the EPC.

\section{Superconductivity} \label{sec:SC}
A Schrieffer-Wolff transformation \cite{Bardeen1955Pines, Schrieffer1966Wolff} yields an effective electron-electron interaction mediated by the phonons. Focusing on zero-momentum pairing, we have $H_{\text{BCS}} = (1/2)\sum_{\boldsymbol{kk}'} \bar{V}_{\boldsymbol{k}'\boldsymbol{k}} d_{\boldsymbol{k}'}^\dagger d_{-\boldsymbol{k}'}^\dagger d_{-\boldsymbol{k}} d_{\boldsymbol{k}}$.
Following the generalized mean-field BCS theory in Refs.~\cite{Maeland2023AprPRL, Maeland2024Thesis, Sigrist}, the linearized FS averaged gap equation is 
\begin{equation}
    \lambda \Delta_{k_\parallel} =  -\frac{N_L S_{\text{FS}}}{N_{\text{samp}}A_{\text{BZ}}} \sum_{k'_\parallel } v_{k'_\parallel}^{-1}  \bar{V}_{k_\parallel k'_\parallel}^{\text{FS}}\Delta_{k'_\parallel },
\end{equation}
where $k_\parallel$ and $k_\perp$ are momenta parallel and perpendicular to the FS, $S_{\text{FS}}$ is the length of the FS, $N_{\text{samp}}$ is the number of evenly spaced points sampled on the FS, $A_{\text{BZ}}$ is the area of the 1BZ, and $v_{k_\parallel} = |\partial\epsilon/\partial k_\perp|$ is the $k_\parallel$-dependent slope of the band perpendicular to the FS. The symmetrized interaction \cite{Sigrist, Maeland2023AprPRL} is $\bar{V}_{\boldsymbol{k}\boldsymbol{k}'} = V_{\boldsymbol{k}\boldsymbol{k}'} - V_{\boldsymbol{k},-\boldsymbol{k}'}$, where, on the FS, $V_{\boldsymbol{k} \boldsymbol{k}'}^{\text{FS}} = -\sum_m g_{\boldsymbol{k}\boldsymbol{k}'}^m g_{-\boldsymbol{k},-\boldsymbol{k}'}^m /\omega_{\boldsymbol{k}-\boldsymbol{k}',m}$. Appendix \ref{app:eeint} discusses some computational details of the electron-electron interaction. The linearized gap equation is an eigenvalue problem, with the eigenvector corresponding to the largest eigenvalue $\lambda$ giving the momentum dependence of the gap function $\Delta_{k_\parallel}$. 
The gap function must be odd in momentum by the Pauli exchange statistics.
The dimensionless coupling $\lambda$ gives an estimate of the critical temperature $T_c$ through $k_B T_c \approx 1.13 \omega_D e^{-1/\lambda}$ with $\omega_D$ being the maximum phonon energy \cite{BCS}.

\begin{figure}
    \centering
    \includegraphics[width=\linewidth]{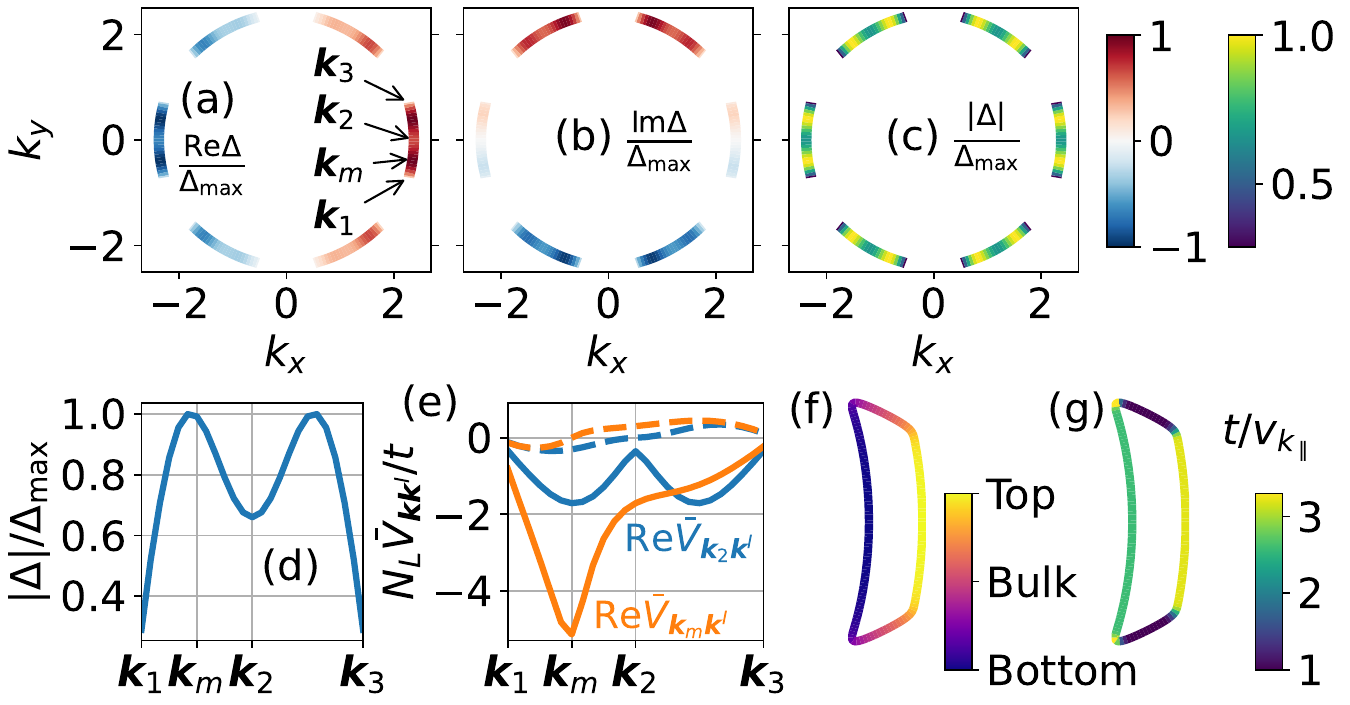}
    \caption{Real (a), imaginary (b), and absolute value (c) of the gap function on the Fermi arc for the bottom surface, which dominates the superconductivity. The unit is the maximum absolute value of the gap $\Delta_{\text{max}}$. The absolute value of the gap is the same on all six arcs, and is shown along one arc in (d). (e) The electron-electron coupling $\bar{V}_{\boldsymbol{k}\boldsymbol{k}'}$ as a function of $\boldsymbol{k}'$ on one arc with $\boldsymbol{k}$ fixed at the center of the arc (blue) and the midpoint between center and endpoint (orange). Solid lines give the real part, while the imaginary part is dashed. Panel (f) repeats the weight of the states on one FS pocket, while (g) shows the inverse slope perpendicular to the FS. Both of these are the same on all six FS pockets. The parameters are $M = 5.6\times 10^{4}/t$, $\chi = 8$, $N_{\text{samp}} = 150$, $L = 40$, and otherwise the same as Figs.~\ref{fig:OBCweight} and \ref{fig:phonon}. Those parameters give $\lambda \approx 0.38$ and $T_c \approx 19.7$~K if $t = 1$~eV.}
    \label{fig:gap}
\end{figure}

We find that the Fermi arc on the bottom surface dominates the superconductivity when $\mu < 0$ at least down to $\mu = -0.1t$. 
Figure~\ref{fig:gap} shows real and imaginary parts of the gap in panels (a) and (b). The gap is $p_x + ip_y$ wave, also called chiral $p$ wave. We find that the chiral $p$-wave nature is insensitive to material parameters and interpret it as coming from SOC through the electronic eigenstates; see Appendix \ref{app:SC}. Interestingly, since the Fermi arc on the bottom surface gives a nondegenerate FS, this is an intrinsic realization of the state proposed by Fu and Kane \cite{Fu2008Kane}. Since TRS is retained \cite{Scheurer2016NCSTSC}, the state falls in symmetry class BDI \cite{Fukui2010FuKaneTopoStability} which has no topologically nontrivial state in 2D \cite{Schnyder2008tenfold}. However, as pointed out in Ref.~\cite{Fu2008Kane}, all that is needed is a very small out-of-plane magnetic field to break the TRS and send the system into class D, realizing a 2D topological superconductor. 
This would yield Majorana bound states in the core of vortices \cite{Read2000MBS},\footnote{By core of vortex we mean the real space location where the gap amplitude goes to zero due to the presence of the vortex.} and a chiral edge state in the case of an odd number of vortices \cite{Bernevig2013}.

The BCS prediction of $T_c$ has an exponential dependence on the material parameters and is usually an overestimate \cite{Allen1975Aug}. 
We have aimed for somewhat realistic parameter values, and so a prediction of $T_c \approx 19.7$~K suggests phonon-mediated surface superconductivity with a measurable $T_c$ is possible in Weyl semimetals. 
The exact behavior of the anisotropic absolute value of the gap is also sensitive to the choice of material parameters. Here, we discuss a peculiarity of phonon-mediated pairing that is not captured by a local Hubbard attraction. As seen in Fig.~\ref{fig:gap}(d), the absolute value of the gap is suppressed in the center of the Fermi arc, where the states are most surface localized. Note that a local Hubbard attraction predicts a maximum gap in the center of the arc \cite{Trama2024TRSWeylSM_SC}. 

The suppression of the gap in the center of the arc makes sense when considering the electron-electron coupling in Fig.~\ref{fig:gap}(e). This shows that the maximum absolute value of $\bar{V}_{\boldsymbol{k}\boldsymbol{k}'}$ as a function of $\boldsymbol{k}'$ occurs when $\boldsymbol{k} = \boldsymbol{k}_m$ at the midpoint between the center of the arc and its endpoint, rather than when $\boldsymbol{k} = \boldsymbol{k}_2$ in the center of the arc. We now argue that this is due to the out-of-plane type EPC. In the center of the arc the electronic states are almost entirely localized on the surface, and so the in-plane type EPC dominates the paring. We see in Fig.~\ref{fig:gap}(e) that the in-plane EPC alone gives a small contribution to the electron-electron coupling, as evidenced by the small value of $\bar{V}_{\boldsymbol{k}_2\boldsymbol{k}_2}$. As we move along the Fermi arc, electronic states start penetrating into the bulk, and the layer closest to the bottom layer also gets a significant occupation. Then, out-of-plane EPC starts contributing. 

For in-plane EPC the factor inside brackets in Eq.~\eqref{eq:gEPC} becomes $(e^{-i\boldsymbol{k}'\cdot \boldsymbol{\delta}}-e^{-i\boldsymbol{k}\cdot\boldsymbol{\delta}})\Bar{\boldsymbol{e}}_{\boldsymbol{k}-\boldsymbol{k}',m}^{z_i}$ and the in-plane EPC behaves like a lattice version of the jellium model \cite{BruusFlensberg}. The in-plane EPC gives zero electron-electron coupling for zero momentum transfer with optical phonons. With acoustic phonons, there is a small nonzero electron-electron coupling in the limit of zero momentum transfer. Meanwhile, in the out-of-plane EPC, $g_{\boldsymbol{k}, \boldsymbol{k}',m}^{\ell\ell' z_i \delta_z}$ does not necessarily go to zero at zero momentum transfer since the phonon eigenstates $\Bar{\boldsymbol{e}}_{\boldsymbol{k}-\boldsymbol{k}',m}^{z_i\pm1}$ and $\Bar{\boldsymbol{e}}_{\boldsymbol{k}-\boldsymbol{k}',m}^{z_i}$ come from different layers. This can give a significant contribution to the electron-electron interaction, especially for optical phonons with a low energy at $\boldsymbol{q} = 0$. To better demonstrate the effect, we tune the phonon parameters to ensure many such modes in Fig.~\ref{fig:phonon}(a). In Appendix \ref{app:SC} we discuss how different phonon parameters influence the quantitative properties of the results.

From the above argument involving out-of-plane EPC, one might wonder why the gap decreases again when the momentum moves toward the edge of the Fermi arc and states become more bulk like, as seen in Fig.~\ref{fig:gap}. By choosing momenta $\boldsymbol{k}$ and $\boldsymbol{k}'$ in different sections of the FS, e.g., bottom Fermi arc and bulk, we find a very low $\bar{V}_{\boldsymbol{k}\boldsymbol{k}'}$. The low coupling is due to the low overlap of the electronic eigenstates. Hence, superconductivity in the top surface Fermi arc, in the bulk. and in the bottom surface Fermi arc nearly decouple. Hence, it is comparable to a three-band superconductor and we should expect one of them to dominate. Therefore, from continuity, the gap on the dominant Fermi arc should decay as the momentum moves towards the bulk like part of the FS. 

Let us also comment on why the bottom surface dominates the superconductivity. In Fig.~\ref{fig:gap}, we show the result when only focusing the the bottom surface Fermi arc, defined as the part of the FS with $W_{\boldsymbol{k}n} < 0.3$. If we do the same for the top surface Fermi arc ($W_{\boldsymbol{k}n} > 0.7$), then we get a similar result though with a lower $\lambda$ and hence a lower $T_c$, see Appendix \ref{app:SC}. In Fig.~\ref{fig:gap}(g), we show the inverse slope on the FS. It is an important part of the DOS and a larger value means the band is less dispersive. We see that the top surface has a flatter band than the bottom surface, while the bottom surface Fermi arc is longer than the top surface Fermi arc. We conjecture that the exact momentum location of the Fermi arc, and hence which electron eigenstates are used influences the coupling strength. See also Refs.~\cite{Brekke2023Dec, Maeland2024Apr, Leraand2025Feb} for discussions of how a varying slope influences the superconducting properties. By comparing Figs.~\ref{fig:gap}(f) and \ref{fig:gap}(g) we see that the bulk like parts have considerably more dispersive bands than the Fermi arcs. Also, the bulk like parts are shorter, together giving a significantly lower bulk DOS compared to the surface. Hence, bulk superconductivity has a much lower $T_c$. 
See Appendix \ref{app:SC} for more details of the competition between surface and bulk superconductivity.

The surface state of a TI is nondegenerate and there is a strong SOC. Therefore, we predict phonons could mediate a spinless chiral $p$-wave pairing there as well. 
However, other studies predict that EPC in a TI is too weak to give a measurable $T_c$ \cite{Parente2013TISCphonon}. The surface state of a TI should not have a significant momentum-dependent penetration into the bulk. Hence, the out-of-plane EPC should be suppressed. We conjecture that the unique property of Weyl semimetals, namely the momentum-dependent bulk penetration of Fermi arc surface states, enables out-of-plane type EPC to enhance the phonon-mediated pairing such that $T_c$ might become measurable.\footnote{Some Dirac semimetals have surface states with properties similar to Weyl semimetals \cite{Gorbar2015MarDirac, Gorbar2015JunDirac, Armitage2018WeylRev, Xu2014DecFermiArcExp, Kargarian2016DiracFermiArc}.}

While we have been motivated by PtBi$_2$, our predictions should be more general and apply to many Weyl semimetals. 
More recent angle-resolved photoemission spectroscopy (ARPES) measurements on PtBi$_2$ indicate that the gap function has nodes in the center of the Fermi arcs \cite{Changdar2025iwave}, in contrast to our prediction. Hence, another mechanism is likely at play, potentially in combination with phonon-mediated pairing. The Weyl semimetals MoTe$_2$ \cite{Naidyuk2018WSC} and TaIrTe$_4$ \cite{Xing2020WSC} appear to be better candidates for dominant phonon-mediated pairing, since EPC is stronger for lighter atoms. 
If an ARPES measurement similar to Ref.~\cite{Changdar2025iwave} reveals a full gap with a suppression in the center of the Fermi arc, then it would be a strong indicator of the mechanism we propose. An accompanying scanning tunneling microscope (STM) observation of zero-bias peaks in the core of vortices, as expected for Majorana bound states, would support the presence of topological superconductivity.
Other signatures of unconventional superconductivity include spontaneous generation of magnetic fields in chiral superconductors \cite{Kvorning2018SpontaneousMag} and features in STM \cite{Sukhachov2023SCandreevSTS, Sukhachov2023SCimpurity, Panigrahi2025SCSTM}.

\section{Conclusion} \label{sec:conclusion}
We show that topological superconductivity can occur intrinsically on the surface of Weyl semimetals due to electron-phonon coupling. 
By treating the system with periodic boundary conditions in two directions and open boundary conditions in one direction we capture two surfaces and the bulk in one formalism. Weak-coupling BCS theory calculations show that superconductivity on the surface dominates. 
We demonstrate that electron-phonon coupling originating with in-plane and out-of-plane hopping behave differently and that the out-of-plane one can be stronger. This gives the strongest coupling for electronic states with a certain degree of penetration into the bulk, such that the maximum absolute value of the gap occurs between the center of the Fermi arc and its endpoints. 

\section*{Acknowledgments}
We thank Jan Carl Budich, Johannes Mitscherling, and Philipp R{\"u}{\ss}mann for useful discussions.
This work was supported by the Deutsche Forschungsgemeinschaft (DFG, German Research Foundation) project SFB 1170 and DFG through the W{\"u}rzburg-Dresden Cluster of Excellence ct.qmat (EXC 2147, project-id 390858490).

\appendix

\section{Electron model} \label{app:Weylmodel}
The electron model with PBC in all three directions is covered in Ref.~\cite{Vocaturo2024PtBi2Effective}. There, the authors show that the model realizes a Weyl semimetal. They also inverse Fourier transform to have OBC in the $z$ direction. Since surface states are our focus, we here provide more details of the treatment of a slab geometry, with PBC in the $xy$ plane and OBC in the $z$ direction.

For the hexagonal crystal we use the basis vectors $\Bar{\boldsymbol{a}}_1 = (0,1,0)$, $\Bar{\boldsymbol{a}}_2 = (\sqrt{3}/2, -1/2, 0)$, and $\Bar{\boldsymbol{a}}_3 = (0,0,1)$.
The nearest-neighbor vectors are
\begin{align}
    &\Bar{\boldsymbol{\delta}}_1 = \frac{1}{2}(\sqrt{3}, 1,0), \Bar{\boldsymbol{\delta}}_2 = (0,1,0), \Bar{\boldsymbol{\delta}}_3 = \frac{1}{2}(-\sqrt{3}, 1,0),\\
    & \Bar{\boldsymbol{\delta}}_4 = \frac{1}{2}(-\sqrt{3}, -1,0),  \Bar{\boldsymbol{\delta}}_5 = (0,-1,0),\\
    &\Bar{\boldsymbol{\delta}}_6 = \frac{1}{2}(\sqrt{3}, -1,0), \Bar{\boldsymbol{\delta}}_7 = (0,0,1), \Bar{\boldsymbol{\delta}}_8 = (0,0,-1).
\end{align}
The 12 next-nearest-neighbor vectors are
\begin{align}
    \Bar{\boldsymbol{\tau}}_{i=\{1,2,3,4,5,6\}} &= \Bar{\boldsymbol{\delta}}_{i=\{1,2,3,4,5,6\}} + (0,0,1), \\
    \Bar{\boldsymbol{\tau}}_{i=\{7,8,9,10,11,12\}} &= \Bar{\boldsymbol{\delta}}_{i=\{1,2,3,4,5,6\}} - (0,0,1).
\end{align}
See an illustration in Fig.~\ref{fig:phonon}(b).

We consider only nearest-neighbor hopping $t_{\ell\ell'}(\Bar{\boldsymbol{\delta}})$. For in-plane hopping we let $t_{AA}(\Bar{\boldsymbol{\delta}}_{i=1,\dots,6}) = t/2, t_{AB}(\Bar{\boldsymbol{\delta}}_{i=1,3,5}) = t_o/2,$ and $ t_{AB}(\Bar{\boldsymbol{\delta}}_{i=2,4,6}) = -t_o/2$. For out-of-plane hopping, $t_{AA}(\Bar{\boldsymbol{\delta}}_{i = 7,8}) = -\beta/2,$ $t_{AB}(\Bar{\boldsymbol{\delta}}_7) = -\beta/2$, and $t_{AB}(\Bar{\boldsymbol{\delta}}_8) = \beta/2$. Finally, $t_{BB}(\Bar{\boldsymbol{\delta}}_i) = -t_{AA}(\Bar{\boldsymbol{\delta}}_i)$ and $t_{BA}(\Bar{\boldsymbol{\delta}}_{i}) = -t_{AB}(\Bar{\boldsymbol{\delta}}_{i})$. 

By defining
\begin{equation}
    f_{\boldsymbol{k}} = \mu_o -t\bqty{\cos (k_y) +2\cos(\frac{\sqrt{3}k_x}{2})\cos(\frac{k_y}{2})},
\end{equation}
\begin{equation}
    g_{\boldsymbol{k}} = t_o \bqty{\sin(k_y)-2 \cos(\frac{\sqrt{3}k_x}{2})\sin(\frac{k_y}{2})},
\end{equation}
and
the vector $\boldsymbol{c}_{\boldsymbol{k}} = (c_{\boldsymbol{k},L,A\uparrow}, c_{\boldsymbol{k},L,A\downarrow}, c_{\boldsymbol{k},L,B\uparrow}, c_{\boldsymbol{k},L,B\downarrow}, $ $c_{\boldsymbol{k},L-1,A\uparrow}, c_{\boldsymbol{k},L-1,A\downarrow}, c_{\boldsymbol{k},L-1,B\uparrow}, c_{\boldsymbol{k},L-1,B\downarrow}, \dots, $ $ c_{\boldsymbol{k},1,A\uparrow}, c_{\boldsymbol{k},1,A\downarrow}, c_{\boldsymbol{k},1,B\uparrow}, c_{\boldsymbol{k},1,B\downarrow})^T$, we can write
\begin{equation}
    H_{\text{el}} = \sum_{\boldsymbol{k}} \boldsymbol{c}_{\boldsymbol{k}}^\dagger H(\boldsymbol{k}) \boldsymbol{c}_{\boldsymbol{k}}.
\end{equation}
Since we do not couple anything beyond nearest-neighbor layers, $H(\boldsymbol{k})$ is block tridiagonal, with $4\times 4$ blocks. The diagonal blocks $H_D(\boldsymbol{k})$ are
\begin{equation}
    H_D(\boldsymbol{k}) = \begin{pmatrix}
        -\mu + f_{\boldsymbol{k}} & 0 & \gamma-ig_{\boldsymbol{k}} & s_{\boldsymbol{k}}  \\
        0 & -\mu + f_{\boldsymbol{k}} & s_{\boldsymbol{k}}^* & \gamma-ig_{\boldsymbol{k}}  \\
        \gamma+ig_{\boldsymbol{k}} & s_{\boldsymbol{k}} & -\mu-f_{\boldsymbol{k}} & 0  \\
        s_{\boldsymbol{k}}^* & \gamma + ig_{\boldsymbol{k}} & 0 & -\mu - f_{\boldsymbol{k}} 
    \end{pmatrix},
\end{equation}
and the upper diagonal blocks $H_U(\boldsymbol{k})$ are
\begin{equation}
    H_U(\boldsymbol{k}) = \frac{1}{2}\begin{pmatrix}
         \beta & 0 & \beta & -s_{\boldsymbol{k}}  \\
         0 &\beta & -s_{\boldsymbol{k}}^* & \beta  \\
         -\beta & -s_{\boldsymbol{k}} & -\beta & 0  \\
         -s_{\boldsymbol{k}}^* & -\beta & 0 & -\beta 
    \end{pmatrix}.
\end{equation}
The lower diagonal blocks $H_L(\boldsymbol{k}) = H_U^\dagger (\boldsymbol{k})$, by Hermiticity.

The high-symmetry points in the 1BZ of the 2D triangular lattice layers are $\boldsymbol{\Gamma} = (0,0)$, $\boldsymbol{M} = (2\pi/\sqrt{3}, 0)$, and $\boldsymbol{K} = (2\pi/\sqrt{3}, 2\pi/3)$.

\section{Phonon model} \label{app:phonon}
\subsection{General framework}
Phonon dispersions can be derived in specific lattice systems using the system's symmetries through a force constant approach \cite{BruusFlensberg, Klogetvedt2023, Syljuasen2024, Thingstad2020Jun, Leraand2025Feb}. There are $dr$ phonon modes, where $d$ is the dimensionality and $r$ is the number of atoms in the basis. $d$ of the phonon modes are acoustic, with zero energy at zero momentum, while the remaining $d(r-1)$ modes are optical, with a nonzero energy  at zero momentum.

Consider a lattice of ions with equilibrium positions $\Bar{\boldsymbol{R}}_{i\alpha}$ and instantaneous positions $\Bar{\boldsymbol{r}}_{i\alpha}(t) = \Bar{\boldsymbol{R}}_{i\alpha} +\Bar{\boldsymbol{u}}_{i\alpha}(t)$. The vector $\Bar{\boldsymbol{u}}_{i\alpha}(t)$ is the displacement from equilibrium. We use $i$ to denote unit cells, and $\alpha = 1,2,\dots,r$ to denote atoms within the basis. The ion Hamiltonian is kinetic plus potential energy. 
We expect small deviations from equilibrium and Taylor expand the potential energy term $V$:
\begin{widetext}
\begin{align}
    \sum_{i\alpha \neq j \beta} V(\Bar{\boldsymbol{r}}_{j\beta}-\Bar{\boldsymbol{r}}_{i\alpha}) &\approx \sum_{i\alpha \neq j \beta} \bigg[ V(\Bar{\boldsymbol{R}}_{j\beta}-\Bar{\boldsymbol{R}}_{i\alpha}) + \sum_\mu \left. \pdv{V}{R_{i\alpha\mu}}\right\rvert_{\text{eq}} u_{i\alpha\mu} +\frac{1}{2} \sum_{\mu\nu} \left.\frac{\partial^2 V}{\partial R_{i\alpha\mu} \partial R_{j\beta\nu}}\right\rvert_{\text{eq}}u_{i\alpha\mu}  u_{j\beta\nu} \bigg]. \label{eq:Vtaylor}
\end{align}
\end{widetext}
The first term is a constant, which we let shift the zero point of energy. The second term acts like a force, which must be zero in equilibrium. The subscript ``eq''~is used to indicate that the derivative is evaluated in equilibrium. We are left with
\begin{align}
    V &= \frac{1}{2}\sum_{i\alpha\mu, j \beta\nu} \left.\frac{\partial^2 V}{\partial R_{i\alpha\mu} \partial R_{j\beta\nu}}\right\rvert_{\text{eq}}u_{i\alpha\mu}  u_{j\beta\nu} \nonumber\\
    &= \frac{1}{2}\sum_{i\alpha\mu, j \beta\nu} \Phi_{\mu\nu}^{\alpha\beta}(\Bar{\boldsymbol{R}}_j-\Bar{\boldsymbol{R}}_i) u_{i\alpha\mu}  u_{j\beta\nu}. \label{eq:Vphi}
\end{align}
$\Phi_{\mu\nu}^{\alpha\beta}(\Bar{\boldsymbol{R}}_j-\Bar{\boldsymbol{R}}_i)$ is the force coefficient on atom $i\alpha$ in the $\mu$ direction, due to the displacement of atom $j\beta$ in the $\nu$ direction.
Note that here $i=j, \alpha=\beta$ is included, which leads to self-force coefficients \cite{Klogetvedt2023}. The stability requirement is used to find the self-force coefficients, $\sum_{j\beta} \Phi_{\mu\nu}^{\alpha\beta}(\Bar{\boldsymbol{R}}_j-\Bar{\boldsymbol{R}}_i) = 0$. It is derived from the fact that the potential should not change if we move all ions by the same arbitrary amount \cite{Syljuasen2024}.

From the symmetry of order of derivatives, we obtain
$\Phi_{\mu\nu}^{\alpha\beta}(\Bar{\boldsymbol{R}}_j-\Bar{\boldsymbol{R}}_i) = \Phi_{\nu\mu}^{\beta\alpha}(\Bar{\boldsymbol{R}}_i-\Bar{\boldsymbol{R}}_j)$.
Let $S$ be a matrix version of a symmetry that leaves the system invariant. We require that also $ \Phi_{\mu\nu}^{\alpha\beta}(\Bar{\boldsymbol{R}}_j-\Bar{\boldsymbol{R}}_i)$ obeys this symmetry, i.e.,
\begin{align}
\label{eq:Phisymm}
     \Phi_{\mu\nu}^{\alpha\beta}(\Bar{\boldsymbol{R}}_j-\Bar{\boldsymbol{R}}_i) &= \sum_{\mu'\nu'} S_{\mu\mu'}^T  \Phi_{\mu'\nu'}^{\alpha\beta}\big(S(\Bar{\boldsymbol{R}}_j-\Bar{\boldsymbol{R}}_i)\big) S_{\nu'\nu} \nonumber\\
    &= [S^T \Phi^{\alpha\beta}\big(S(\Bar{\boldsymbol{R}}_j-\Bar{\boldsymbol{R}}_i)\big) S]_{\mu\nu}.
\end{align}
From these symmetries, we can limit the number of free parameters.

From Newton's second law, we get the equation of motion
\begin{equation}
\label{eq:eqm}
    M_\alpha \partial^2_t u_{i\alpha\mu} = -\pdv{V}{u_{i\alpha\mu}} = -\sum_{j\beta\nu}  \Phi_{\mu\nu}^{\alpha\beta}(\Bar{\boldsymbol{R}}_j-\Bar{\boldsymbol{R}}_i)  u_{j\beta\nu},
\end{equation}
where $M_\alpha$ is the mass of atom $\alpha$ in the unit cell.
The harmonic nature of the equation motivates a plane-wave ansatz
\begin{equation}
\label{eq:deviationPhononOp}
    u_{i\alpha\mu}(t) = \sum_{\Bar{\boldsymbol{q}}} \frac{1}{\sqrt{M_\alpha}} c_{\alpha\mu}(\Bar{\boldsymbol{q}}) e^{i(\Bar{\boldsymbol{q}}\cdot \Bar{\boldsymbol{R}}_{i\alpha}-\omega t)},
\end{equation}
where $c_{\alpha\mu}(\Bar{\boldsymbol{q}})$ is an expansion coefficient, and $\omega$ is a momentum-dependent frequency we find by inserting the ansatz into the equation of motion. The result is
$
    \omega^2 c_{\alpha\mu}(\Bar{\boldsymbol{q}}) = \sum_{\beta\nu} D_{\mu\nu}^{\alpha\beta}(\Bar{\boldsymbol{q}})c_{\beta\nu}(\Bar{\boldsymbol{q}}),
$
where we defined
\begin{equation}
\label{eq:Dmatrix}
    D_{\mu\nu}^{\alpha\beta}(\Bar{\boldsymbol{q}}) = \sum_j \frac{1}{\sqrt{M_\alpha M_\beta}}\Phi_{\mu\nu}^{\alpha\beta}(\Bar{\boldsymbol{R}}_j-\Bar{\boldsymbol{R}}_i)e^{i\Bar{\boldsymbol{q}}\cdot (\Bar{\boldsymbol{R}}_{j\beta}-\Bar{\boldsymbol{R}}_{i\alpha})}.
\end{equation}
The $dr \times dr$ dynamical matrix $D(\Bar{\boldsymbol{q}})$ has a set of eigenvalues $\omega_{\Bar{\boldsymbol{q}}m}^2$. Their roots are the phonon spectrum $\omega_{\Bar{\boldsymbol{q}}m}$ and their corresponding normalized eigenvectors are $\hat{e}_{\Bar{\boldsymbol{q}}m}$.

We can quantize the Hamiltonian in terms of bosonic phonon operators as \cite{Klogetvedt2023, Syljuasen2024}
\begin{equation}
    u_{i\alpha\mu}(t) = \sum_{\Bar{\boldsymbol{q}}m} \frac{1}{\sqrt{2NM_\alpha \omega_{\Bar{\boldsymbol{q}}m}}} e_{\Bar{\boldsymbol{q}}m}^{\alpha\mu} [a_{-\Bar{\boldsymbol{q}},m}^\dagger(t)+a_{\Bar{\boldsymbol{q}}m}(t)]e^{i\Bar{\boldsymbol{q}}\cdot \Bar{\boldsymbol{R}}_{i\alpha}},
\end{equation}
where the time-dependence of the phonon operator is $a_{\Bar{\boldsymbol{q}}m}(t) = a_{\Bar{\boldsymbol{q}}m}e^{-i\omega_{\Bar{\boldsymbol{q}}m}t}$ and $N$ is the total number of unit cells.
Then, the phonon Hamiltonian is
$
    H_{\text{ph}} = \sum_{\Bar{\boldsymbol{q}}m} \omega_{\Bar{\boldsymbol{q}}m} a_{\Bar{\boldsymbol{q}}m}^\dagger a_{\Bar{\boldsymbol{q}}m}.
$

\subsection{Hexagonal lattice}
In our case, we consider a 3D crystal with a one-atomic basis ($r=1$ and $\alpha,\beta$ indices become superfluous) with symmetries given by the P31m space group. There are three mirror planes meeting at $120^\circ$ normal to the $xy$ plane, with the intersection being the $z$ axis. Also, there is a three-fold rotational symmetry about the $z$ axis. 

\begin{table*}[th]
    \centering
    \caption{Parameterization of $\Phi_{\mu\nu}(\Bar{\boldsymbol{R}}_{ji})/M$ for nearest and next-nearest neighbors. The self-force coefficients are $\Phi_{xx}(0)/M = \Phi_{yy}(0)/M = -6\gamma_1-2\gamma_3-6\rho_1-6\rho_2$, $\Phi_{zz}(0)/M = -6\gamma_4-2\gamma_5-6\rho_5-6\rho_6$, and $\Phi_{\mu\nu}(0)/M = 0$ if $\mu\neq \nu$.}
    \label{tab:force}
    \begin{ruledtabular}
    \begin{tabular}{cccccccccccccc}
        $\mu\nu\diagdown\Bar{\boldsymbol{R}}_{ji}$ & $\Bar{\boldsymbol{\delta}}_{1}$ & $\Bar{\boldsymbol{\delta}}_{2}$ & $\Bar{\boldsymbol{\delta}}_{3}$ & $\Bar{\boldsymbol{\delta}}_{4}$ & $\Bar{\boldsymbol{\delta}}_{5}$ & $\Bar{\boldsymbol{\delta}}_{6}$ & $\Bar{\boldsymbol{\delta}}_{i=\{7,8\}}$ & $\Bar{\boldsymbol{\tau}}_{i=\{1,10\}}$ & $\Bar{\boldsymbol{\tau}}_{i=\{2,11\}}$ & $\Bar{\boldsymbol{\tau}}_{i=\{3,12\}}$ & $\Bar{\boldsymbol{\tau}}_{i=\{4,7\}}$ & $\Bar{\boldsymbol{\tau}}_{i=\{5,8\}}$ & $\Bar{\boldsymbol{\tau}}_{i=\{6,9\}}$ \\[1mm]
        \hline
        $xx$ & $\gamma_1$ & $\gamma_1$ & $\gamma_1$ & $\gamma_1$ & $\gamma_1$ & $\gamma_1$ & $\gamma_3$ & $\rho_1$ & $\rho_2$ & $\rho_1$ & $\rho_2$ & $\rho_1$ & $\rho_2$  \\
        $yy$ & $\gamma_1$ & $\gamma_1$ & $\gamma_1$ & $\gamma_1$ & $\gamma_1$ & $\gamma_1$ & $\gamma_3$ & $\rho_1$ & $\rho_2$ & $\rho_1$ & $\rho_2$ & $\rho_1$ & $\rho_2$  \\
        $zz$ & $\gamma_4$ & $\gamma_4$ & $\gamma_4$ & $\gamma_4$ & $\gamma_4$ & $\gamma_4$ & $\gamma_5$ & $\rho_5$ & $\rho_6$ & $\rho_5$ & $\rho_6$ & $\rho_5$ & $\rho_6$  \\
        $xy$ & 0 & 0 & 0 & 0 & 0 & 0 & 0 & 0 & 0 & 0 & 0 & 0 & 0 \\
        $yx$ & 0 & 0 & 0 & 0 & 0 & 0 & 0 & 0 & 0 & 0 & 0 & 0 & 0 \\
        $xz$ & $\gamma_6$ & 0 & $-\gamma_6$ & $\gamma_7$ & 0 & $-\gamma_7$ & 0 & $\rho_7$ & 0 & $-\rho_7$ & $\rho_8$ & 0 & $-\rho_8$ \\
        $zx$ & $\gamma_7$ & 0 & $-\gamma_7$ & $\gamma_6$ & 0 & $-\gamma_6$ & 0 & $\rho_7$ & 0 & $-\rho_7$ & $\rho_8$ & 0 & $-\rho_8$ \\
        $yz$ & $\frac{\gamma_6}{\sqrt{3}}$ & $\frac{-2\gamma_7}{\sqrt{3}}$ & $\frac{\gamma_6}{\sqrt{3}}$ & $\frac{\gamma_7}{\sqrt{3}}$  & $\frac{-2\gamma_6}{\sqrt{3}}$ & $\frac{\gamma_7}{\sqrt{3}}$ & 0 & $\frac{\rho_7}{\sqrt{3}}$ & $\frac{-2\rho_8}{\sqrt{3}}$ & $\frac{\rho_7}{\sqrt{3}}$ & $\frac{\rho_8}{\sqrt{3}}$ & $\frac{-2\rho_7}{\sqrt{3}}$ & $\frac{\rho_8}{\sqrt{3}}$ \\
        $zy$ & $\frac{\gamma_7}{\sqrt{3}}$ & $\frac{-2\gamma_6}{\sqrt{3}}$ & $\frac{\gamma_7}{\sqrt{3}}$ & $\frac{\gamma_6}{\sqrt{3}}$  & $\frac{-2\gamma_7}{\sqrt{3}}$ & $\frac{\gamma_6}{\sqrt{3}}$ & 0 & $\frac{\rho_7}{\sqrt{3}}$ & $\frac{-2\rho_8}{\sqrt{3}}$ & $\frac{\rho_7}{\sqrt{3}}$ & $\frac{\rho_8}{\sqrt{3}}$ & $\frac{-2\rho_7}{\sqrt{3}}$ & $\frac{\rho_8}{\sqrt{3}}$ \\
    \end{tabular}
    \end{ruledtabular}
\end{table*}

For rotations around the $z$ axis, we have the two relevant rotations
\begin{equation}
    R_{120} = \begin{pmatrix}
        -1/2 & -\sqrt{3}/2 & 0 \\
        \sqrt{3}/2 & -1/2 & 0 \\
        0 & 0 & 1
    \end{pmatrix},~R_{240} = \begin{pmatrix}
        -1/2 & \sqrt{3}/2 & 0 \\
        -\sqrt{3}/2 & -1/2 & 0 \\
        0 & 0 & 1
    \end{pmatrix}.
\end{equation}
We have one mirror symmetry about the $yz$ plane represented by $M_x$. 
Furthermore, there are mirror symmetries along the two lines $k_y = \pm k_x/\sqrt{3}$, named $M_1$ and $M_2$, respectively. We find
\begin{align}
    &M_x = \begin{pmatrix}
        -1 & 0 & 0 \\
        0 & 1 & 0 \\
        0 & 0 & 1
    \end{pmatrix},~
    M_1 = \begin{pmatrix}
        1/2 & \sqrt{3}/2 & 0 \\
        \sqrt{3}/2 & -1/2 & 0 \\
        0 & 0 & 1
    \end{pmatrix},\nonumber\\
    &M_2 = \begin{pmatrix}
        1/2 & -\sqrt{3}/2 & 0 \\
        -\sqrt{3}/2 & -1/2 & 0 \\
        0 & 0 & 1
    \end{pmatrix}.
\end{align}

The quantity $\Bar{\boldsymbol{R}}_j-\Bar{\boldsymbol{R}}_i = \Bar{\boldsymbol{R}}_{ji}$ in $\Phi_{\mu\nu}(\Bar{\boldsymbol{R}}_{ji})$ is summed over all inter-unit-cell distances. We limit the description up to next-nearest neighbors. Using the symmetries, one can reduce the number of independent parameters. 
In the interest of brevity we leave the detailed symmetry consideration for the interested reader, and simply give one example and state the final result.

Using Eq.~\eqref{eq:Phisymm}, we have for mirror $yz$:
\begin{align}
    &\Phi(\Bar{\boldsymbol{R}}_{ji}) = M_x^T \Phi(M_x \Bar{\boldsymbol{R}}_{ji}) M_x \nonumber\\
    &= \begin{pmatrix}
        \Phi_{xx}(M_x \Bar{\boldsymbol{R}}_{ji}) & -\Phi_{xy}(M_x \Bar{\boldsymbol{R}}_{ji}) & -\Phi_{xz}(M_x \Bar{\boldsymbol{R}}_{ji}) \\
        -\Phi_{yx}(M_x \Bar{\boldsymbol{R}}_{ji}) & \Phi_{yy}(M_x \Bar{\boldsymbol{R}}_{ji}) & \Phi_{yz}(M_x \Bar{\boldsymbol{R}}_{ji}) \\
        -\Phi_{zx}(M_x \Bar{\boldsymbol{R}}_{ji}) & \Phi_{zy}(M_x \Bar{\boldsymbol{R}}_{ji}) & \Phi_{zz}(M_x \Bar{\boldsymbol{R}}_{ji})
    \end{pmatrix}.
\end{align}
This means that, e.g., $\Phi_{xy}(\Bar{\boldsymbol{R}}_{ji}) = 0$ for all $\Bar{\boldsymbol{R}}_{ji}$ with $x$ component 0. Table \ref{tab:force} lists the result of the full symmetry analysis.
There are 12 free parameters $\gamma_1, \gamma_3, \gamma_4,$ $ \gamma_5, \gamma_6, \gamma_7, $ $\rho_1, \rho_2, \rho_5,$ $ \rho_6, \rho_7, \rho_8$ with unit energy squared. We further simplify to $\rho_i = \rho = -\gamma_1/10\sqrt{2}$, $\gamma_4 = \gamma_1, \gamma_5 = \gamma_3, \gamma_7 = \gamma_6$ giving only three parameters $\gamma_1, \gamma_3, \gamma_6$ that we then treat as phenomenological and set to values to get a reasonable range of phonon energies \cite{Bashlakov2022PtBi2phononExp, Benedek2010PhononOBC}, with the constraint that the dynamical matrix $D(\Bar{\boldsymbol{q}})$ is positive semidefinite at all $\Bar{\boldsymbol{q}}$ so that the crystal is stable \cite{Enzner2025phonon}. 

The elements of the dynamical matrix $D(\Bar{\boldsymbol{q}})$ can now be found straightforwardly from Eq.~\eqref{eq:Dmatrix}. 
The three bulk acoustic modes are found from this matrix and shown in Fig.~\ref{fig:phonon}(a) at $q_z = 0$ along high symmetry lines. Now let us consider the correction from having OBC in the $z$ direction.

\subsection{Phonon spectrum with one open boundary condition}
See Refs.~\cite{Lucas1968PhononOBC, Benedek2010PhononOBC} for other studies of phonons with open boundary conditions, which includes examples of the range of phonon energies for certain materials. Reference~\cite{Benedek2010PhononOBC} also discusses surface phonons and enhanced EPC at the surface. The origin is that interatomic distances may change close to the surface due to the interface with vacuum. Here, we ignore any changes in interatomic distances, and focus simply on the changes from having OBC. 
We must return to the equation of motion, and now insert an ansatz that assumes only the 2D component of momentum is well defined, while $z_i$ denotes layer indices. In a way, a slab geometry is like treating the system as having $r = L$ sublattices. So, we expect $3L$ phonon modes, of which 3 are acoustic and $3(L-1)$ are optical modes (here in the sense of having nonzero energy at zero in-plane momentum). 

Now introduce a plane-wave ansatz only for $\boldsymbol{q} = (q_x, q_y)$ while keeping the $z_i$ dependence,
\begin{equation}
    u_{i\alpha\mu}(t) = u_{x_i y_i z_i \alpha\mu}(t) = \sum_{\boldsymbol{q}} \frac{1}{\sqrt{M_\alpha}} c_{z_i\alpha\mu}(\boldsymbol{q}) e^{i(\boldsymbol{q} \cdot  \boldsymbol{R}_{i\alpha}-\omega t)}.
\end{equation}
Inserting into Eq.~\eqref{eq:eqm}, gives
\begin{align}
    \omega^2 c_{z_i\alpha\mu}(\boldsymbol{q}) =& \sum_{z_j\beta\nu} D_{\mu\nu}^{z_i\alpha z_j\beta}(\boldsymbol{q})c_{z_j\beta\nu}(\boldsymbol{q}).
\end{align}
The phonon spectrum $\omega_{\boldsymbol{q}m}$ is the square roots of the eigenvalues of the $drL \times drL$ matrix,
\begin{equation}
    D_{\mu\nu}^{z_i\alpha z_j\beta}(\boldsymbol{q}) = \sum_{x_j y_j} \frac{1}{\sqrt{M_\alpha M_\beta}}\Phi_{\mu\nu}^{\alpha\beta}(\Bar{\boldsymbol{R}}_j-\Bar{\boldsymbol{R}}_i)e^{i\boldsymbol{q}\cdot (\boldsymbol{R}_{j\beta}-\boldsymbol{R}_{i\alpha}) }.
\end{equation}
The corresponding eigenvector elements are $e_{\boldsymbol{q}m}^{z_i \alpha \mu}$.
We quantize the displacements as
\begin{equation}
\label{eq:uphononOBC}
    u_{i\alpha\mu} = \sum_{\boldsymbol{q} m} \frac{1}{\sqrt{2N_L M_\alpha \omega_{m}(\boldsymbol{q})}} e_{\boldsymbol{q}m}^{z_i \alpha \mu} (a_{-\boldsymbol{q},m}^\dagger+a_{\boldsymbol{q}m})e^{i\boldsymbol{q}\cdot \boldsymbol{R}_{i\alpha}},
\end{equation}
where $N_L$ is the number of sites per layer.

For the hexagonal crystal, the elements of the dynamical matrix are now
\begin{widetext}
\begin{align}
    D_{xx}^{z_i z_i}(\boldsymbol{q}) =&  -6\gamma_1-2\gamma_3-6\rho_1-6\rho_2 +2\gamma_1\bqty{2\cos(\frac{\sqrt{3}q_x}{2})\cos(\frac{q_y}{2}) +\cos q_y }.
\end{align}
\begin{align}
    D_{xx}^{z_i,z_i+1}(\boldsymbol{q}) =&  \gamma_3 + \rho_1\pqty{e^{i\pqty{\frac{\sqrt{3}q_x}{2}+\frac{q_y}{2}}}+e^{-i\pqty{\frac{\sqrt{3}q_x}{2}-\frac{q_y}{2}}}+e^{-iq_y}}  + \rho_2 \pqty{e^{-i\pqty{\frac{\sqrt{3}q_x}{2}+\frac{q_y}{2}}}+e^{i\pqty{\frac{\sqrt{3}q_x}{2}-\frac{q_y}{2}}}+e^{iq_y}}. 
\end{align}
$D_{xx}^{z_i,z_i-1}(\boldsymbol{q}) = [D_{xx}^{z_i,z_i+1}(\boldsymbol{q})]^*$, $D_{yy}^{z_i,z_i}(\boldsymbol{q}) = D_{xx}^{z_i,z_i}(\boldsymbol{q})$, $D_{yy}^{z_i,z_i\pm1}(\boldsymbol{q}) = D_{xx}^{z_i,z_i\pm1}(\boldsymbol{q})$.

\begin{align}
    D_{zz}^{z_i z_i}(\boldsymbol{q}) =&  -6\gamma_4-2\gamma_5-6\rho_5-6\rho_6 +2\gamma_4\bqty{2\cos(\frac{\sqrt{3}q_x}{2})\cos(\frac{q_y}{2}) +\cos q_y }.
\end{align}
\begin{align}
    D_{zz}^{z_i,z_i+1}(\boldsymbol{q}) =&  \gamma_5 + \rho_5\pqty{e^{i\pqty{\frac{\sqrt{3}q_x}{2}+\frac{q_y}{2}}}+e^{-i\pqty{\frac{\sqrt{3}q_x}{2}-\frac{q_y}{2}}}+e^{-iq_y}}+ \rho_6 \pqty{e^{-i\pqty{\frac{\sqrt{3}q_x}{2}+\frac{q_y}{2}}}+e^{i\pqty{\frac{\sqrt{3}q_x}{2}-\frac{q_y}{2}}}+e^{iq_y}}.
\end{align}
$D_{zz}^{z_i,z_i-1}(\boldsymbol{q}) = [D_{zz}^{z_i,z_i+1}(\boldsymbol{q})]^*$, $D_{xy}^{z_i z_i}(\boldsymbol{q}) = D_{xy}^{z_i,z_i\pm1}(\boldsymbol{q}) = 0$, $D_{yx}^{z_i,z_j}(\boldsymbol{q}) = D_{xy}^{z_i,z_j}(\boldsymbol{q})$.

\begin{align}
    D_{xz}^{z_i z_i}(\boldsymbol{q}) =&  \gamma_6 \pqty{e^{i\pqty{\frac{\sqrt{3}q_x}{2}+\frac{q_y}{2}}}-e^{-i\pqty{\frac{\sqrt{3}q_x}{2}-\frac{q_y}{2}}}}+\gamma_7 \pqty{e^{-i\pqty{\frac{\sqrt{3}q_x}{2}+\frac{q_y}{2}}}-e^{i\pqty{\frac{\sqrt{3}q_x}{2}-\frac{q_y}{2}}}}.
\end{align}
\begin{align}
    D_{xz}^{z_i,z_i+1}(\boldsymbol{q}) =&  \rho_7\pqty{e^{i\pqty{\frac{\sqrt{3}q_x}{2}+\frac{q_y}{2}}}-e^{-i\pqty{\frac{\sqrt{3}q_x}{2}-\frac{q_y}{2}}}}  + \rho_8 \pqty{e^{-i\pqty{\frac{\sqrt{3}q_x}{2}+\frac{q_y}{2}}}-e^{i\pqty{\frac{\sqrt{3}q_x}{2}-\frac{q_y}{2}}}}. 
\end{align}
$D_{xz}^{z_i,z_i-1}(\boldsymbol{q}) = [D_{xz}^{z_i,z_i+1}(\boldsymbol{q})]^*$, $D_{zx}^{z_i,z_i}(\boldsymbol{q}) = [D_{xz}^{z_i,z_i}(\boldsymbol{q})]^*$, $D_{zx}^{z_i,z_i+1}(\boldsymbol{q}) = D_{xz}^{z_i,z_i+1}(\boldsymbol{q})$, $D_{zx}^{z_i,z_i-1}(\boldsymbol{q}) = [D_{xz}^{z_i,z_i+1}(\boldsymbol{q})]^*$.

\begin{align}
    D_{yz}^{z_i, z_i}(\boldsymbol{q}) =&  \frac{\gamma_6}{\sqrt{3}} \pqty{e^{i\pqty{\frac{\sqrt{3}q_x}{2}+\frac{q_y}{2}}}+e^{-i\pqty{\frac{\sqrt{3}q_x}{2}-\frac{q_y}{2}}}-2e^{-iq_y}}  +\frac{\gamma_7}{\sqrt{3}}\pqty{e^{-i\pqty{\frac{\sqrt{3}q_x}{2}+\frac{q_y}{2}}}+e^{i\pqty{\frac{\sqrt{3}q_x}{2}-\frac{q_y}{2}}}-2e^{iq_y}}. 
\end{align}
\begin{align}
    D_{yz}^{z_i,z_i+1}(\boldsymbol{q}) =&  \frac{\rho_7}{\sqrt{3}} \pqty{e^{i\pqty{\frac{\sqrt{3}q_x}{2}+\frac{q_y}{2}}}+e^{-i\pqty{\frac{\sqrt{3}q_x}{2}-\frac{q_y}{2}}}-2e^{-iq_y}}  +\frac{\rho_8}{\sqrt{3}}\pqty{e^{-i\pqty{\frac{\sqrt{3}q_x}{2}+\frac{q_y}{2}}}+e^{i\pqty{\frac{\sqrt{3}q_x}{2}-\frac{q_y}{2}}}-2e^{iq_y}}.
\end{align}
$D_{yz}^{z_i,z_i-1}(\boldsymbol{q}) = [D_{yz}^{z_i,z_i+1}(\boldsymbol{q})]^*$, $D_{zy}^{z_i, z_i}(\boldsymbol{q}) = [D_{yz}^{z_i, z_i}(\boldsymbol{q})]^*$, $D_{zy}^{z_i,z_i+1}(\boldsymbol{q}) = D_{yz}^{z_i,z_i+1}(\boldsymbol{q})$, $D_{zy}^{z_i,z_i-1}(\boldsymbol{q}) = [D_{yz}^{z_i,z_i+1}(\boldsymbol{q})]^*$.
\end{widetext}

The dynamical matrix is tridiagonal with $3\times 3$ blocks. When $z_i = 1$ and $z_i = L$ the stability criterion changes since there is no layer below and above, respectively. In both cases $-2\gamma_3-6\rho_1-6\rho_2 \to -\gamma_3-3\rho_1-3\rho_2$ in $D_{xx}^{z_i z_i}(\boldsymbol{q})$ and $D_{yy}^{z_i z_i}(\boldsymbol{q})$, while in $D_{zz}^{z_i z_i}(\boldsymbol{q})$, $-2\gamma_5-6\rho_5-6\rho_6 \to -\gamma_5-3\rho_5-3\rho_6$.

The phonon Hamiltonian in the slab geometry is
$
    H = \sum_{\boldsymbol{q} m} \omega_{\boldsymbol{q}m} a_{\boldsymbol{q}m}^\dagger a_{\boldsymbol{q}m},
$
where $m$ now runs over $3L$ phonon modes. The phonon spectrum in the slab geometry is shown as black lines in Fig.~\ref{fig:phonon}(a).

\section{Electron-phonon coupling} \label{app:EPC}
We derive the EPC by Taylor expanding the instantaneous hopping term around the equilibrium positions \cite{Thingstad2020Jun, Leraand2025Feb},
\begin{equation}
    t_{\ell\ell'}(\Bar{\boldsymbol{r}}_{ij}) \approx t_{\ell\ell'}(\Bar{\boldsymbol{R}}_{ij}) + \Bar{\boldsymbol{u}}_{ij} \cdot \left.\nabla_{\Bar{\boldsymbol{r}}_{ij}} t_{\ell\ell'}(\Bar{\boldsymbol{r}}_{ij})\right\vert_{\Bar{\boldsymbol{r}}_{ij} = \Bar{\boldsymbol{R}}_{ij}},
\end{equation}
where $\Bar{\boldsymbol{r}}_{ij} = \Bar{\boldsymbol{r}}_{i}-\Bar{\boldsymbol{r}}_{j}$, $\Bar{\boldsymbol{u}}_{ij} = \Bar{\boldsymbol{r}}_{ij}-\Bar{\boldsymbol{R}}_{ij}$, and $\left.\nabla_{\Bar{\boldsymbol{r}}_{ij}} t_{\ell\ell'}(\Bar{\boldsymbol{r}}_{ij})\right\vert_{\Bar{\boldsymbol{r}}_{ij} = \Bar{\boldsymbol{R}}_{ij}}$ denotes the derivative of $t_{\ell\ell'}(\Bar{\boldsymbol{r}}_{ij})$ with respect to $\Bar{\boldsymbol{r}}_{ij}$ evaluated at equilibrium. 
Initially we have a hopping term,
\begin{equation}
    H_{\text{hop}} = -\sum_{\langle i,j \rangle, \ell, \ell', \sigma} t_{\ell\ell'}(\Bar{\boldsymbol{r}}_i-\Bar{\boldsymbol{r}}_j) c_{i\ell\sigma}^\dagger c_{j\ell' \sigma}.
\end{equation}
We then keep only the equilibrium part in the hopping term,
\begin{equation}
    H_{\text{hop}} = -\sum_{\langle i,j \rangle, \ell, \ell', \sigma} t_{\ell\ell'}(\Bar{\boldsymbol{R}}_i-\Bar{\boldsymbol{R}}_j) c_{i\ell\sigma}^\dagger c_{j\ell' \sigma},
\end{equation}
while the deviations are treated as EPC:
\begin{equation}
    H_{\text{EPC}} = -\sum_{\langle i,j \rangle, \ell, \ell', \sigma} (\Bar{\boldsymbol{u}}_i-\Bar{\boldsymbol{u}}_j) \cdot \left.\nabla_{\Bar{\boldsymbol{r}}_{ij}} t_{\ell\ell'}(\Bar{\boldsymbol{r}}_{ij})\right\vert_{\Bar{\boldsymbol{r}}_{ij} = \Bar{\boldsymbol{R}}_{ij}} c_{i\ell\sigma}^\dagger c_{j\ell' \sigma}.
\end{equation}
We rewrite the derivative as $\left.\nabla_{\Bar{\boldsymbol{r}}_{ij}} t_{\ell\ell'}(\Bar{\boldsymbol{r}}_{ij})\right\vert_{\Bar{\boldsymbol{r}}_{ij} = \Bar{\boldsymbol{R}}_{ij}} \equiv \nabla_{\Bar{\boldsymbol{\delta}}}t_{\ell\ell'}(\Bar{\boldsymbol{\delta}})$,
and let $i \to i+\delta$, $j \to i$.
Then,
\begin{align}
    H_{\text{EPC}} &= -\sum_{i, \Bar{\boldsymbol{\delta}}, \ell, \ell', \sigma} (\Bar{\boldsymbol{u}}_{i+\delta}-\Bar{\boldsymbol{u}}_i) \cdot \nabla_{\Bar{\boldsymbol{\delta}}}t_{\ell\ell'}(\Bar{\boldsymbol{\delta}}) c_{i+\delta,\ell\sigma}^\dagger c_{i\ell' \sigma}.
\end{align}
It is possible to model atomic orbitals, calculate $t_{\ell\ell'}(\Bar{\boldsymbol{r}}_i-\Bar{\boldsymbol{r}}_j)$ as an overlap integral, and then compute the derivatives $\nabla_{\Bar{\boldsymbol{\delta}}}t_{\ell\ell'}(\Bar{\boldsymbol{\delta}})$ \cite{Leraand2025Feb}. 
We instead keep the orbitals unspecified and let $\nabla_{\Bar{\boldsymbol{\delta}}}t_{\ell\ell'}(\Bar{\boldsymbol{\delta}}) = -\chi \Bar{\boldsymbol{\delta}} t_{\ell\ell'}(\Bar{\boldsymbol{\delta}})$ \cite{Thingstad2020Jun}. 
As shown in Ref.~\cite{Leraand2025Feb}, $\chi$ is inversely proportional to the spread of the atomic orbitals. Given that we limit the description to nearest-neighbor hopping, the atomic orbitals should have a small spread, and so we choose a larger value of $\chi$ compared to Ref.~\cite{Thingstad2020Jun}. Alternatively, if we were to use a smaller value of $\chi$, then longer ranged hopping would also contribute to the EPC such that we could still get a significant coupling \cite{Leraand2025Feb}.

We insert the phonon quantization of ionic displacements from Eq.~\eqref{eq:uphononOBC} and Fourier transform to get
\begin{align}
\label{eq:HEPCc}
    H_{\text{EPC}} &= \sum_{\boldsymbol{k} z_i \boldsymbol{q}m} \sum_{\ell, \ell', \sigma} \sum_{\delta_z} g_{\boldsymbol{k}+\boldsymbol{q}, \boldsymbol{k}, m}^{\ell\ell' z_i \delta_z}  (a_{-\boldsymbol{q},m}^\dagger+a_{\boldsymbol{q}m})\nonumber\\
    &\times c_{\boldsymbol{k}+\boldsymbol{q},z_i+\delta_z,\ell\sigma}^\dagger c_{\boldsymbol{k},z_i,\ell' \sigma},
\end{align}
where $g_{\boldsymbol{k}+\boldsymbol{q}, \boldsymbol{k}, m}^{\ell\ell' z_i \delta_z}$ is defined in Eq.~\eqref{eq:gEPC}.
Note that unlike a local Hubbard interaction, we now have interlayer coupling when $\delta_z = \pm1$. 
Furthermore, note that the sum over $\delta_z$ includes all eight nearest-neighbor vectors, also the six in-plane ones with $\delta_z = 0$. We transform the expression in Eq.~\eqref{eq:HEPCc} to the band basis to get the EPC Hamiltonian in Eq.~\eqref{eq:HEPC}.

\section{Electron-electron interaction} \label{app:eeint}
We refer the reader to Refs.~\cite{Bardeen1955Pines, Schrieffer1966Wolff, Maeland2024Thesis} for details of how to perform the Schrieffer-Wolff transformation. Note that we perform the Schrieffer-Wolff transformation and the subsequent BCS mean-field theory in the band basis. Focusing on the nondegenerate band with a FS, we have
\begin{equation}
    H_{\text{BCS}} = \sum_{\boldsymbol{k}\boldsymbol{k}'} V_{\boldsymbol{k}'\boldsymbol{k}} d_{\boldsymbol{k}'}^\dagger d_{-\boldsymbol{k}'}^\dagger d_{-\boldsymbol{k}} d_{\boldsymbol{k}},
\end{equation}
with 
\begin{equation}
    V_{\boldsymbol{k}'\boldsymbol{k}} = \sum_m \frac{g_{\boldsymbol{k}'\boldsymbol{k}}^m g_{-\boldsymbol{k}',-\boldsymbol{k}}^m \omega_{\boldsymbol{k}'-\boldsymbol{k},m}}{(\epsilon_{\boldsymbol{k}}-\epsilon_{\boldsymbol{k}'})^2-\omega_{\boldsymbol{k}'-\boldsymbol{k},m}^2}.
\end{equation}
Note that $V_{-\boldsymbol{k}',-\boldsymbol{k}} = V_{\boldsymbol{k}'\boldsymbol{k}}$, so the symmetrized interaction \cite{Sigrist, Maeland2023AprPRL} is $\bar{V}_{\boldsymbol{k}'\boldsymbol{k}} = V_{\boldsymbol{k}'\boldsymbol{k}} - V_{\boldsymbol{k}',-\boldsymbol{k}}$.

\begin{figure*}[th]
    \centering
    \includegraphics[width=0.9\linewidth]{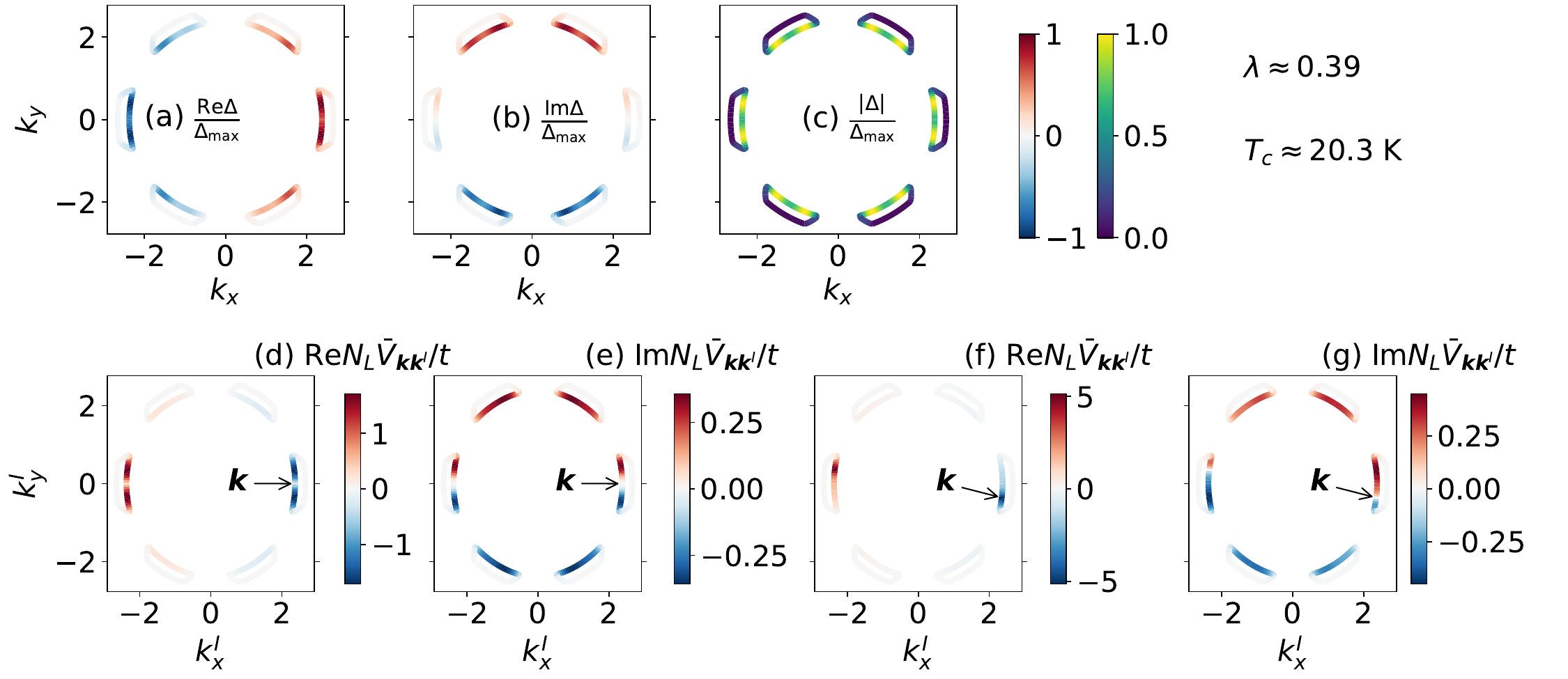}
    \caption{Real (a), imaginary (b), and absolute value (c) of the gap function on the full Fermi surface. The values are scaled by the maximum absolute value of the gap $\Delta_{\text{max}}$. The absolute value of the gap is the same on all six pockets. Panels (d)-(g) give the electron-electron coupling $\bar{V}_{\boldsymbol{k}\boldsymbol{k}'}$ as a function of $\boldsymbol{k}'$ with $\boldsymbol{k}$ fixed as indicated. Note that the coupling is stronger at $\boldsymbol{k}' = \boldsymbol{k}$ in (f) than in (d) which results in a suppression of the absolute value of the gap in the center of the arc. The parameters are $N_{\text{samp}} = 330$ and otherwise the same as in Fig.~\ref{fig:gap}. Specifically, $t_o/t = 1.5$, $\beta/t = -1.5$,  $\mu/t = -0.05, \mu_o/t = 0.2, \alpha/t = -0.18$, $\gamma/t = -0.2$, $\gamma_1 = -(0.005t)^2$, $\gamma_3 = 0.45\gamma_1$, $\gamma_6 = 1.5 \gamma_1$, $M = 5.6\times 10^{4}/t$, $\chi = 8$, and $L=40$. To get $T_c$ in kelvin, we assume $t = 1$~eV. }
    \label{fig:gapFullFS}
\end{figure*}

\subsection{Gauge dependence}
Since it is $\bar{V}_{\boldsymbol{k}\boldsymbol{k}'}^{\text{FS}} = -\sum_m g_{\boldsymbol{k}\boldsymbol{k}'}^m g_{-\boldsymbol{k},-\boldsymbol{k}'}^m /\omega_{\boldsymbol{k}-\boldsymbol{k}',m} + \sum_m g_{\boldsymbol{k},-\boldsymbol{k}'}^m g_{-\boldsymbol{k},\boldsymbol{k}'}^m /\omega_{\boldsymbol{k}+\boldsymbol{k}',m}$ which enters in the gap equation we focus on its behavior. The combinations $g_{\boldsymbol{k}\boldsymbol{k}'}^m g_{-\boldsymbol{k},-\boldsymbol{k}'}^m$ and $g_{\boldsymbol{k},-\boldsymbol{k}'}^m g_{-\boldsymbol{k},\boldsymbol{k}'}^m$ both contain products $e_{\boldsymbol{q}m}^{z_i\mu} e_{-\boldsymbol{q},m}^{z'_{i}\nu}$. The phonon eigenvectors have the property $\hat{e}_{-\boldsymbol{q},m} = [\hat{e}_{\boldsymbol{q}m}]^*$ which can be shown from the fact that the displacement vector is real \cite{Klogetvedt2023}. Hence, the electron-electron interaction is independent of any complex local gauge factor in the numerically obtained phonon eigenvectors.

The electron transformation coefficients are more challenging. As a reminder, $g_{\boldsymbol{k}, \boldsymbol{k}'}^{m} = \sum_{\ell\ell'\sigma z_i \delta_z} g_{\boldsymbol{k}, \boldsymbol{k}',m}^{\ell\ell' z_i \delta_z}v_{\boldsymbol{k},z_i+\delta_z,\ell\sigma}^* v_{\boldsymbol{k}',z_i,\ell'\sigma},$
where $c_{\boldsymbol{k},z_i,\ell\sigma} = v_{\boldsymbol{k},z_i,\ell\sigma}d_{\boldsymbol{k}}$. Here, when $\boldsymbol{k} \neq \boldsymbol{k}'$, local gauge factors of the type $e^{if(\boldsymbol{k})}$ in $v_{\boldsymbol{k},z_i,\ell\sigma}$ do not cancel. We find that random local gauges are problematic and yield a discontinuous $\bar{V}_{\boldsymbol{k}\boldsymbol{k}'}^{\text{FS}}$. From the physical picture that $\bar{V}_{\boldsymbol{k}\boldsymbol{k}'}^{\text{FS}}$ should be continuous, we argue that a specific global gauge must be chosen. Inspired by the typical form of analytic eigenvectors we choose to set a specific element of all eigenvectors to be real and positive. 

Still, a gauge dependence carries over into the electron-electron interaction. This is not dramatic, the important thing is that any measurable quantity remains gauge independent. This detail was already pointed out in Ref.~\cite{Scheurer2016NCSTSC}. In our case we find that the gap in the band basis can be either $p_x+ip_y$ or $p_x-ip_y$ depending on if the eigenvector element that is chosen real and positive is associated with spin up or down. In principle, once an eigenvector is found, it can be multiplied by $(k_x \pm i k_y)/\sqrt{k_x^2+k_y^2}$ and remain an eigenvector. Then, the gap in the band basis will be $f_x \pm i f_y$. The absolute value of the gap on the other hand, an experimentally accessible quantity, remains unchanged by how the global gauge is chosen. The same applies to the topological classification and any other property with measurable consequences. Alternatively, the gap can be transformed back to the original basis to eliminate the gauge dependence, see below. In the results we show, the eigenvector element related to bottom surface $z_i = 1$, orbital $\ell  = B$, and spin $\sigma = \uparrow$ is set real and positive. 

\subsection{Zero momentum transfer}
The three phonon modes with zero energy at $\boldsymbol{q} = 0$ in the slab geometry have eigenvectors with elements $1/L$ at the $x,y$, or $z$ position at each layer and otherwise zero. Hence, they correspond to moving all ions the same amount in $x,y,$ or $z$ direction. So, both for the in-plane and out-of-plane type EPC, the numerator in $V_{\boldsymbol{k}\boldsymbol{k}'}^{\text{FS}}$ goes to zero when the denominator goes to zero. Therefore, the value at zero momentum transfer for acoustic phonons is defined by limits, again using the physically reasonable constraint that $\bar{V}_{\boldsymbol{k}\boldsymbol{k}'}^{\text{FS}}$ should be continuous on the FS. We define the value at $k'_\parallel = k_\parallel$ as $V_{k_\parallel k_\parallel} = (V_{k_\parallel, k_\parallel+a_\parallel}+ V_{k_\parallel, k_\parallel-a_\parallel}) /2$ where $a_\parallel$ is a very short arc along the FS.

\section{Superconductivity} \label{app:SC}
\subsection{Bulk and surface separability and origin of gap symmetry}
Figures \ref{fig:gapFullFS}(d)-\ref{fig:gapFullFS}(g) show $\bar{V}_{\boldsymbol{k}\boldsymbol{k}'}^{\text{FS}}$ as a function of $\boldsymbol{k}'$ on the full FS with $\boldsymbol{k}$ fixed at the center of the Fermi arc on bottom surface and halfway between the center and endpoint of the bottom surface Fermi arc. 
In both cases, we see that the coupling is negligible for $\boldsymbol{k}'$ in the bulk and top surface parts of the FS, supporting the claim in Sec.~\ref{sec:SC} that top surface, bulk, and bottom surface superconductivity nearly decouple. We also see a clearly chiral $p$-wave form of the coupling. 
It has a form similar to $\bar{V}_{\boldsymbol{k}\boldsymbol{k}'}^{\text{FS}} \sim -(k_x+ik_y)(k'_x-ik'_y)$ though it is significantly more anisotropic. As mentioned in Sec.~\ref{sec:Electrons}, the SOC factor $s_{\boldsymbol{k}}$ also takes a chiral $p$-wave form. The chiral $p$-wave form of $\bar{V}_{\boldsymbol{k}\boldsymbol{k}'}^{\text{FS}}$ originates with SOC through the electron transformation coefficients. In normal metals, phonons mediate $s$-wave spin singlet pairing through a coupling which to a simplest approximation is a negative constant. Here we capture the anisotropies in the EPC. Working in the band basis, the electron-electron interaction also features electron transformation coefficients that enter in an odd in momentum combination. In a way, we have the anisotropic $s$-wave pairing from phonons multiplied by chiral $p$-wave from SOC resulting in an odd-parity gap function as required for a nondegenerate FS.

In the present model, the occupation on the layer closest to the surface is not negligible, even in the center of the arc. Hence, the out-of-plane type EPC also contributes there. One can imagine a situation where the surface states have an almost vanishing occupation on the layer closest to the surface in the center of the Fermi arcs. We have found such a behavior in the diamond lattice model of Ref.~\cite{Okugawa2014FermiArcEffective}. In that case, we expect the suppression of the absolute value of the gap in the center of the Fermi arc can become more pronounced.

\begin{figure*}[th]
    \centering
    \includegraphics[width=0.9\linewidth]{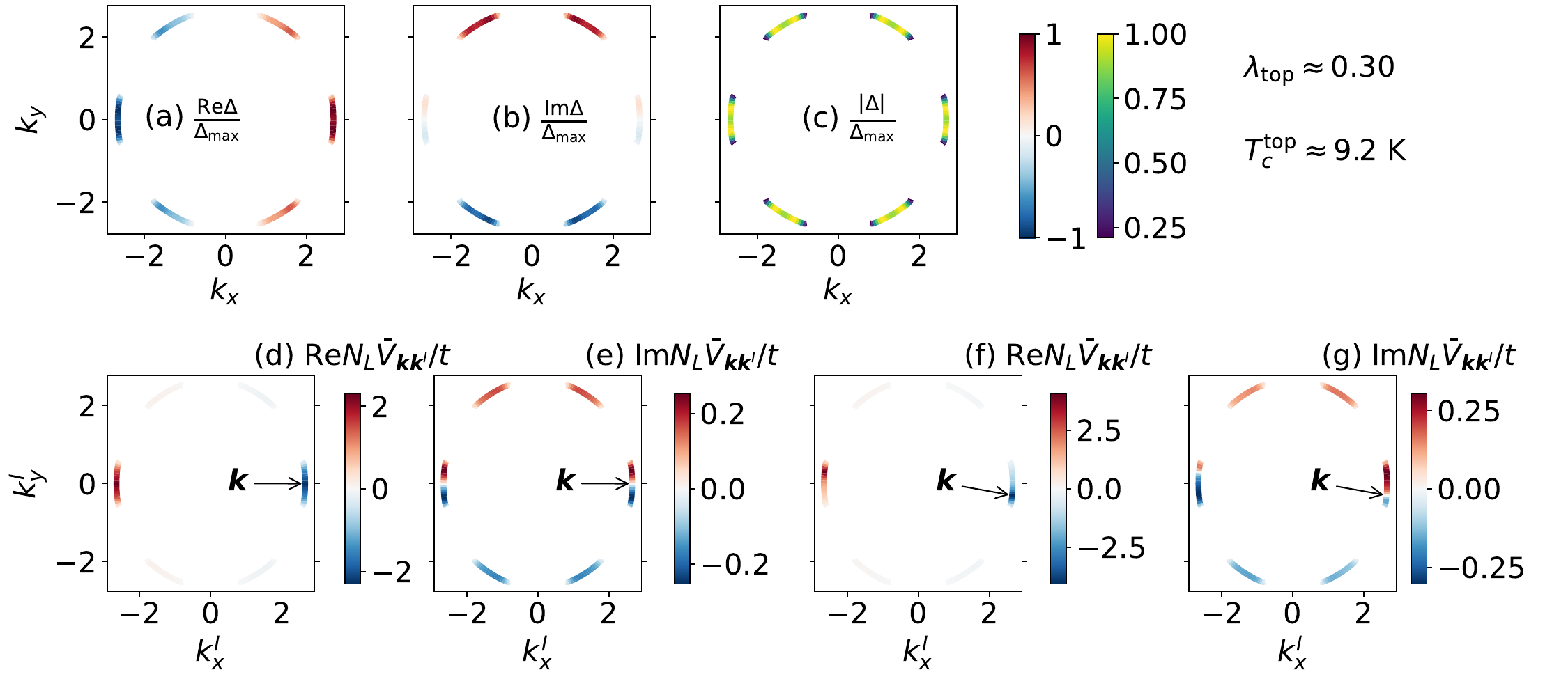}
    \caption{Real (a), imaginary (b), and absolute value (c) of the gap function on the top surface Fermi arc. The values are scaled by the maximum absolute value of the gap $\Delta_{\text{max}}$. The absolute value of the gap is the same on all six arcs. Panels (d)-(g) give the electron-electron coupling $\bar{V}_{\boldsymbol{k}\boldsymbol{k}'}$ as a function of $\boldsymbol{k}'$ with $\boldsymbol{k}$ fixed as indicated. Note again that the coupling is stronger at $\boldsymbol{k}' = \boldsymbol{k}$ in (f) than in (d) which results in a suppression of the absolute value of the gap in the center of the arc. However, the difference is smaller than on the bottom surface, leading to a weaker suppression. The parameters are $N_{\text{samp}} = 126$ and otherwise the same as Fig.~\ref{fig:gapFullFS}. In this figure only, the eigenvector element related to top surface $z_i = L$, orbital $\ell  = A$, and spin $\sigma = \uparrow$ is set real and positive. }
    \label{fig:gaptop}
\end{figure*}

\subsection{Superconducting gap in original basis}
Reference~\cite{Vocaturo2024PtBi2Effective} performed a symmetry analysis in the spin basis of possible superconducting pairing in Fermi arcs in the model we have adopted. We can transform our gap function to the original basis and compare, using $d_{\boldsymbol{k}}^\dagger = \sum_{z_i \ell \sigma} v_{\boldsymbol{k}z_i \ell\sigma} c_{\boldsymbol{k}z_i \ell\sigma}^\dagger$. Consider the following term in the Hamiltonian
\begin{align}
    &\sum_{\boldsymbol{k}} \Delta_{\boldsymbol{k}} d_{\boldsymbol{k}}^\dagger d_{-\boldsymbol{k}}^\dagger \nonumber\\
    &=\sum_{\boldsymbol{k}}\sum_{\substack{z_i\ell\sigma\\z'_i \ell'\sigma'}} \Delta_{\boldsymbol{k}} v_{\boldsymbol{k}z_i \ell\sigma} v_{-\boldsymbol{k},z'_i,\ell',\sigma'}  c_{\boldsymbol{k}z_i \ell\sigma}^\dagger c_{-\boldsymbol{k},z'_i,\ell',\sigma'}^\dagger  \nonumber\\
    &=\sum_{\boldsymbol{k}}\sum_{\substack{z_i\ell,\sigma\\z'_i \ell'\sigma'}} \Delta_{\boldsymbol{k}z_i z'_i \ell \ell' \sigma \sigma'} c_{\boldsymbol{k}z_i \ell\sigma}^\dagger c_{-\boldsymbol{k},z'_i,\ell',\sigma'}^\dagger.
\end{align}
This highlights why it is so convenient to work in the band basis where we have a single gap function $\Delta_{\boldsymbol{k}}$. In the original basis there are $16 L^2$ gap functions $\Delta_{\boldsymbol{k}z_i z'_i \ell \ell' \sigma \sigma'}$, which we find numerically from the gap in the band basis, using the eigenvectors of the electron Hamiltonian. The momentum dependence of the gap functions are in principle infinite series of lattice harmonics, but we read off the symmetry by counting nodes. Let us focus on the $16$ gap functions with $z_i = z'_i = 1$ on the bottom surface. We just mention that those with $z_i = 1(2)$ and $z'_i = 2(1)$ are only slightly smaller and odd in layer indices. The gaps are even in the orbital indices. For all combinations of orbitals we find $s$-wave gaps for spin singlet and chiral $p$-wave gaps for $\sigma\sigma' = \uparrow\uparrow$ and $\sigma\sigma' = \downarrow\downarrow$ in a time-reversal symmetric combination \cite{Maeland2023AprPRL, Sigrist}. This corresponds well with the symmetry analysis in Ref.~\cite{Vocaturo2024PtBi2Effective}, suggesting that phonons are the likely pairing mechanism behind such symmetries of the gap function. The symmetry analysis in Ref.~\cite{Vocaturo2024PtBi2Effective} also predicted a coexisting mixed-spin spin-triplet $f$-wave gap. Its amplitude is negligible in our calculation.

\subsection{Competition of surface and bulk superconductivity}\label{sec:surfacebulkSC}
Figures \ref{fig:gapFullFS}(a)-\ref{fig:gapFullFS}(c) show the superconducting gap function on the full FS. The gap on the bottom surface is the same as shown in Fig.~\ref{fig:gap} when considering only the bottom surface Fermi arc. Meanwhile, the gap in the bulk and top surface parts of the FS are negligible and decrease as $L$ increases. Hence, there will be a range of temperatures where only a gap on one surface is measurable. It is worth noting however, that the gap in the bulk and top surface parts of the FS are not exactly zero. This is quite typical for weakly coupled gaps, and can also be seen in, e.g., three-band superconductors \cite{Aase2023TRSB}. Hence, one could talk about an unmeasurable \cite{OLeary2025PtBi2, Kuibarov2025ARPES} bulk superconductivity that exists purely due to a weak coupling to the dominant surface superconductivity. We expect the small discrepancy in $\lambda$ and $T_c$ compared to only treating the bottom surface Fermi arc (Fig.~\ref{fig:gap}) will decrease when increasing $L$. It is worth noting that the calculation in Sec.~\ref{sec:SC} including only the bottom surface Fermi arc, indicates that surface superconductivity can arise independently of the bulk \cite{Nomani2023FermiArcSCDOS}.

In Fig.~\ref{fig:gaptop}, we show the solution of the gap equation when considering only the top surface Fermi arc. With the same parameters as when considering bottom surface Fermi arc and full FS, we get $\lambda_{\text{top}} \approx 0.30$ and $T_c^{\text{top}} \approx 9.2$~K. The gap still has a suppression in the center of the arc, even though $\bar{V}_{\boldsymbol{k}\boldsymbol{k}'}$ is not as suppressed for $\boldsymbol{k} = \boldsymbol{k}'$ in the center of the arc as for the bottom surface. We would expect the gap at the top surface to start increasing more rapidly at $T_c^{\text{top}}$ and coexist with a larger gap on the bottom surface. Below $T_c^{\text{top}}$ but above the bulk $T_c^{\text{bulk}}$ the top and bottom surface are then practically independent 2D systems, each realizing a 2D topological superconductor when applying a small out-of-plane magnetic field. An increasing sample thickness ($L$) will make the surfaces more and more independent of each other.

As expected from Ref.~\cite{Nomani2023FermiArcSCDOS}, the surface superconductivity has a weak dependence on the chemical potential. We find that bulk and top surface have the same $T_c$ at $\mu = 0$ and that the top surface superconductivity dominates at $\mu > 0$. The bulk superconductivity naturally has a stronger dependence on $\mu$, as there is no bulk FS at $\mu = 0$. We can also make a prediction of the bulk $T_c$ from the model with a slab geometry, by only considering the part of the FS which is bulk like, $0.3 < W_{\boldsymbol{k}n} < 0.7$. Interestingly, the prediction of bulk $T_c$ decreases as we increase $L$. We interpret this as an enhancement of bulk superconductivity coming from the existence of the surfaces, even when the top and bottom Fermi arcs are ignored. At $L = 40$ we find $\lambda_{\text{bulk}} \approx 0.11$ giving $T_c^{\text{bulk}} \approx 0.04$~K with the same parameters used in Fig.~\ref{fig:gapFullFS}. One could also study bulk superconductivity by having PBC in all three directions. That would eliminate the surfaces, and we believe the prediction for $T_c$ would then be even lower, as that is essentially the $L \to \infty$ limit. The exact value of $|\mu|$ where bulk superconductivity would dominate over surface superconductivity is $L$ dependent but $|\mu|$ larger than 50~meV, as in the experiment on PtBi$_2$ \cite{Kuibarov2024FermiArcSCNat}, is reasonable within our choice of material parameters. With parameters as in Fig.~\ref{fig:gapFullFS}, we find that $\mu = -0.12t$ gives approximately the same $T_c$ for bottom surface Fermi arc and bulk like parts of the FS when considered separately, suggesting that bulk superconductivity will dominate at $|\mu| > 0.12t$.

The above general predictions about top surface, bulk, and bottom surface superconductivity appear to be in good agreement with the experiment \cite{Kuibarov2024FermiArcSCNat} where the two surfaces show a different $T_c$ ($\approx\! 14$~K and $\approx\! 8$~K) both of which are higher than the bulk $T_c$ ($\approx\! 0.6$~K) measured in the material earlier \cite{Shipunov2020ExpWeylSC}. 

From solving linearized BCS-type gap equations, the DOS on the Fermi level is a relatively straightforward by-product \cite{Maeland2024Thesis}. In the slab geometry, we consider a local DOS with a layer index dependence \cite{Hodt2024Aug}, $N_F(z_i) = \sum_{\boldsymbol{k}\ell\sigma}|v_{\boldsymbol{k}z_i \ell\sigma}|^2 \delta(\epsilon_{\boldsymbol{k}})$. We sum it over all the layers, and use that at each $\boldsymbol{k}$, $\sum_{z_i\ell\sigma} |v_{\boldsymbol{k}z_i \ell\sigma}|^2 = 1$ to get
\begin{equation}
    N_F = \sum_{z_i}N_F(z_i) =\sum_{\boldsymbol{k}}\delta(\epsilon_{\boldsymbol{k}}) = \frac{N_L S_{\text{FS}}}{N_{\text{samp}}A_{\text{BZ}}} \sum_{k'_\parallel } \abs{\pdv{\epsilon}{k'_\perp}}^{-1}.
\end{equation}
We find that for the full FS, $tN_F/N_L \approx 1.07$, for the top surface Fermi arc, $tN_F^{\text{top}}/N_L \approx 0.49$, for the bulk, $tN_F^{\text{bulk}}/N_L \approx 0.07$, and for the bottom surface Fermi arc, $tN_F^{\text{bottom}}/N_L \approx 0.53$ with the same parameters as Fig.~\ref{fig:gapFullFS} except $N_{\text{samp}}$ which varies depending on which part of the FS is included. We interpret $tN_F/N_L$ as the dimensionless DOS per site in a layer, summed over all the layers. Since the DOS is an important part of the dimensionless coupling $\lambda$, the differences in $T_c$ between the bottom surface, top surface, and bulk can largely be understood from the differences in DOS.

\subsection{Consequence of breaking time-reversal symmetry}
It is worth noting that in a Weyl semimetal with broken inversion symmetry, additionally breaking time-reversal symmetry means there is no guarantee that for a Fermi arc at $\boldsymbol{k}$ there is a corresponding one at $-\boldsymbol{k}$. This scenario is disadvantageous for superconductivity. However, we only require a very weak out-of-plane magnetic field, and assume its effect on the location of Fermi arcs is small compared to the bandwidth of the phonons, such that the superconducting properties should remain the same as in the zero-field calculation. In fact, superconductivity in PtBi$_2$ has been observed to survive quite strong magnetic fields \cite{Bashlakov2022PtBi2phononExp, Moreno2025PtBi2vortex}. A Weyl semimetal with broken time-reversal symmetry and retained inversion symmetry would also be an exciting system for the mechanism we propose. Then, class D topological superconductivity mediated by phonons should appear directly on the surfaces, without the need for a magnetic field. This assumes a type of inversion symmetric Weyl semimetal where a Fermi arc at $\boldsymbol{k}$ has a partner at $-\boldsymbol{k}$. Since inversion symmetry is broken at the surface, that may require further symmetries to be present \cite{TopoSCrevSato}.

\subsection{Sensitivity to phonon properties}

\begin{figure}[b]
    \centering
    \includegraphics[width=\linewidth]{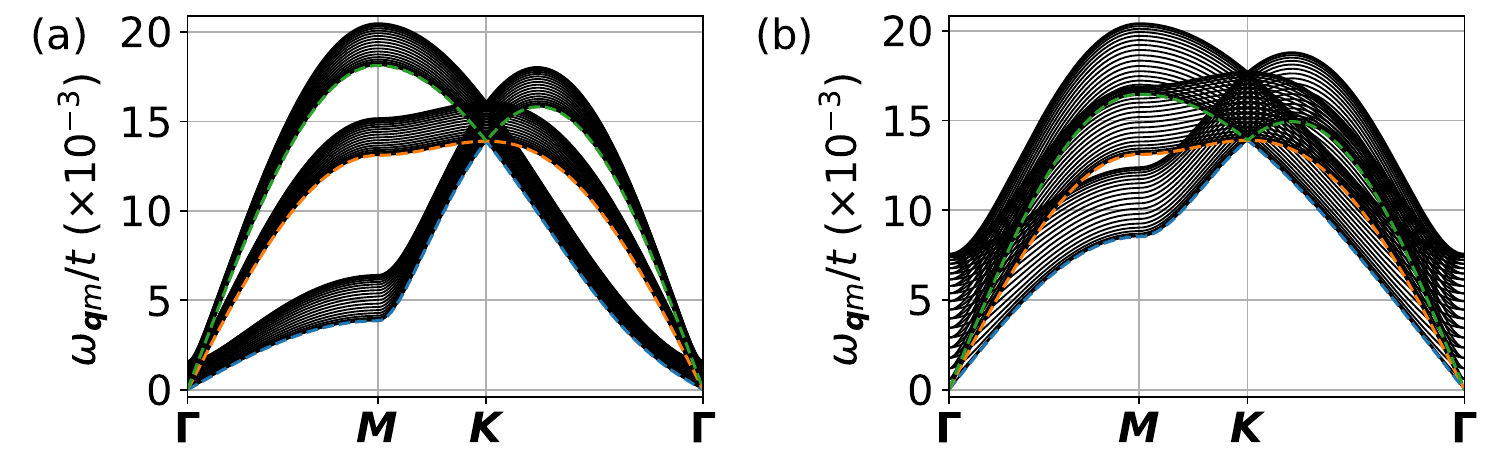}
    \caption{ Black lines show phonon modes in the slab geometry. Colored lines show the three modes with PBC in all directions. (a) Phonon parameters as in Fig.~\ref{fig:phonon}: $\gamma_1 = -(0.005t)^2$, $\gamma_3 = 0.45\gamma_1$, $\gamma_6 = 1.5 \gamma_1$. (b) $\gamma_1 = -(0.005t)^2$, $\gamma_3 = \gamma_6 = \gamma_1$. In both panels $L = 20$. }
    \label{fig:phononcompare}
\end{figure}

\begin{figure*}[thb]
    \centering
    \includegraphics[width=0.9\linewidth]{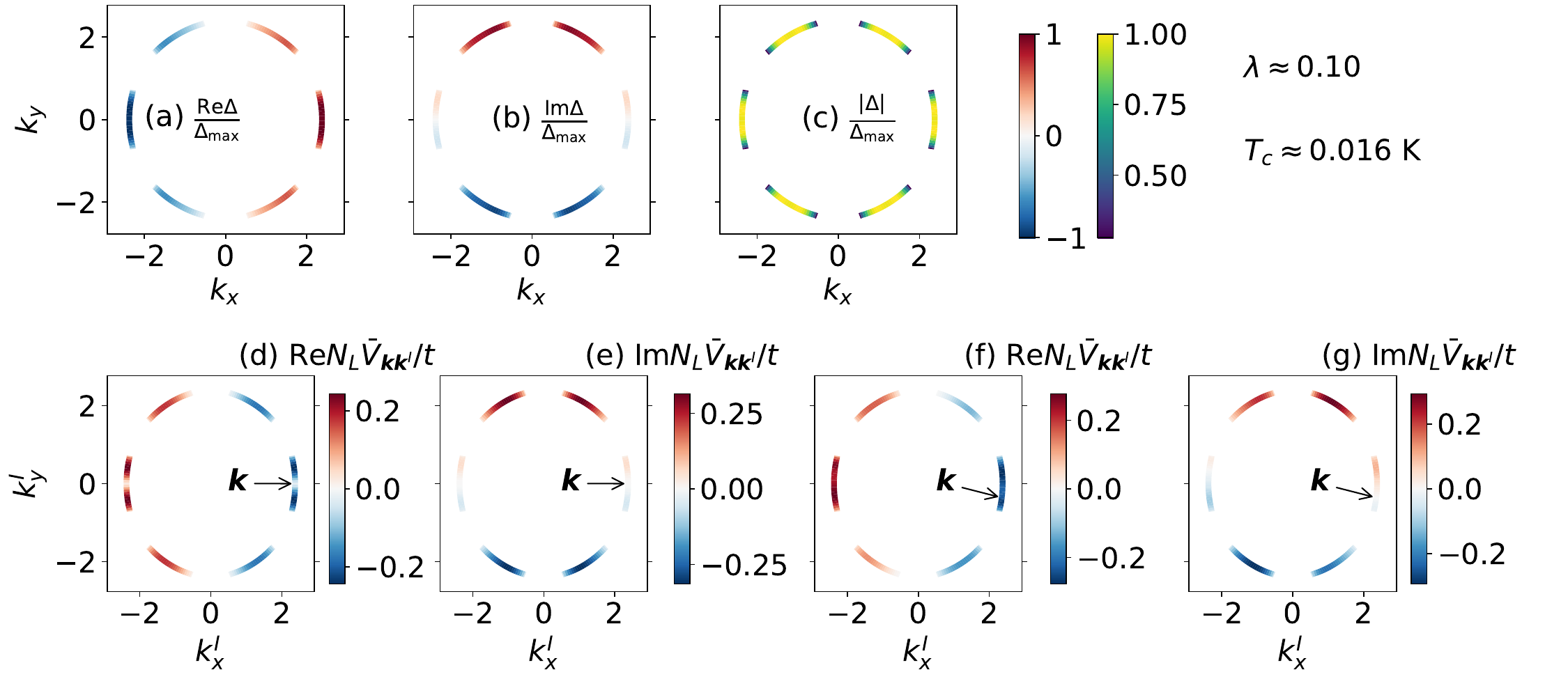}
    \caption{Real (a), imaginary (b), and absolute value (c) of the gap function on the bottom surface Fermi arc. The values are scaled by the maximum absolute value of the gap $\Delta_{\text{max}}$. The absolute value of the gap is the same on all six arcs. Panels (d)-(g) give the electron-electron coupling $\bar{V}_{\boldsymbol{k}\boldsymbol{k}'}$ as a function of $\boldsymbol{k}'$ with $\boldsymbol{k}$ fixed as indicated. The parameters are $N_{\text{samp}} = 150$, $\gamma_3 = \gamma_6 = \gamma_1$ and otherwise the same as Fig.~\ref{fig:gapFullFS}. }
    \label{fig:gapbottompheq}
\end{figure*}

Here we show that the main results are robust towards changing the phonon properties. In Fig.~\ref{fig:phononcompare}, we compare the phonon spectrum with two different sets of phonon parameters, namely, those used in the rest of the work to emphasize the effect of the suppression in the center of the arc, and other parameters where the optical phonons have a greater energy spread at $\boldsymbol{q} = 0$.  
Figure \ref{fig:gapbottompheq} shows the gap on the bottom surface Fermi arc with the new phonon properties. The same general features are present, namely, a chiral $p$-wave gap function and dominant bottom surface superconductivity. There is still a suppression of the absolute value of the gap in the center of the arc, though now the suppression is only about 2\%. The dimensionless coupling $\lambda \approx 0.10$ results in $T_c \approx 0.016$~K. As expected, the prediction of $T_c$ depends on the choice of material parameters. Since $\lambda \sim \chi^2/M$ we could increase $\lambda$ by considering lighter atoms. Additionally, a reduction of the phonon energy range by reducing $|\gamma_1|$ would increase $\lambda$ and hence $T_c$. 
For instance, with $\gamma_1 = -(0.0037t)^2$, $M = 4\times10^4/t$ and otherwise the same parameters as Fig.~\ref{fig:gapbottompheq}, we get $\omega_D/t \approx 0.015$, $\lambda \approx 0.26$, and $T_c \approx 4.4$~K.

\bibliography{main.bbl}

\newpage
\onecolumngrid

\phantomsection
\begin{large}
\begin{center}
    \textbf{Erratum: Phonon-mediated intrinsic topological superconductivity in Fermi arcs [Phys. Rev. B 112, 104507 (2025)]} \label{sec:erratum}
\end{center}
\end{large}

\renewcommand{\theequation}{ERR\arabic{equation}}
\renewcommand{\theHequation}{ERR\arabic{equation}}
\setcounter{equation}{0}

\let\oldsection\section
\renewcommand{\section}[1]{\hspace*{1em}\textit{#1}.}

\section{Introduction}
In the original paper \cite{Maeland2025Jun}, all results relying on the numerical computation of the phonon-mediated electron-electron interaction contain an error. Here we present corrected versions of these results. The main conclusion, namely that phonon-mediated pairing yields intrinsic topological superconductivity with a spinless chiral $p$-wave gap function on the surface of Weyl semimetals, remains valid. 
Furthermore, there is an error in Table \ref{tab:force} of the original paper regarding the phonon model. A typo in the symmetry analysis led to us setting groups of two force constants equal as enforced by symmetry. In fact, there is no symmetry argument that forces them to be equal. The original results then simply present the case where they are chosen to be equal. We present an updated version of Table \ref{tab:force} showing general force constants.

We made a point in the original paper that the phonon-mediated interaction typically prefers a local minimum of the gap magnitude in the center of the arc, unlike the prediction from a local Hubbard attraction \cite{Trama2024TRSWeylSM_SC}. This claim was repeated in the abstract, introduction, and conclusion. Here, we show that the typical case in fact is that a phonon-mediated attraction prefers a maximum gap magnitude in the center of the Fermi arcs. We find a new parameter set where a phonon-mediated attraction still gives a local minimum in the center of the arc. This involves introducing a dependence on the orbital degree of freedom also in the out-of-plane hopping. The mechanism behind the local minimum in the center of the Fermi arcs remains the same as stated in the original paper, but with more stringent requirements on the electronic surface states.

The motivation to focus on the local minimum of the gap in the center of the Fermi arcs was the observation of $i$-wave pairing in Ref.~\cite{Changdar2025iwave}. A local minimum in  the gap amplitude appeared to indicate that phonon-mediated pairing might be relevant there despite the different gap symmetry. In Ref.~\cite{Maeland2026Jul}, the authors show that phonon-mediated pairing combined with Coulomb repulsion can explain the $i$-wave gap when the surface state bandwidth is comparable to the maximum phonon energy. The mechanism also works when phonon-mediated pairing alone predicts maximum gap magnitude in the center of the arc. The unique feature of the phonon-mediated attraction in the slab geometry, compared to a local Hubbard attraction, is that it has a long-ranged component in real space. When Coulomb repulsion then prevents short-ranged real-space Cooper pairing, the long-ranged attraction from phonons can generate higher angular momentum pairing such as $i$ wave. In momentum space, we identify the long-ranged component of the phonon-mediated interaction as a peak in the interaction at zero momentum transfer.

\begin{table}[ht]
\centering
\caption{Parameterization of $\Phi_{\mu\nu}(\Bar{\boldsymbol{R}}_{ji})/M$ for nearest neighbors.
The self-force coefficients are $\Phi_{xx}(0)/M=\Phi_{yy}(0)/M=-3\gamma_{1}-3\gamma_{2}-2\gamma_{3}-3\rho_{1}-3\rho_{2}-3\rho_{3}-3\rho_{4}$,
$\Phi_{zz}(0)/M=-6\gamma_{4}-2\gamma_{5}-6\rho_{5}-6\rho_{6}$, and $\Phi_{\mu\nu}(0)/M=0$ if $\mu\neq\nu$.}
\label{tab:phi_delta}
\begin{ruledtabular}
\begin{tabular}{cccccccc}
$\mu\nu\diagdown\Bar{\boldsymbol{R}}_{ji}$ & $\Bar{\boldsymbol{\delta}}_{1}$ & $\Bar{\boldsymbol{\delta}}_{2}$ & $\Bar{\boldsymbol{\delta}}_{3}$ & $\Bar{\boldsymbol{\delta}}_{4}$ & $\Bar{\boldsymbol{\delta}}_{5}$ & $\Bar{\boldsymbol{\delta}}_{6}$ & $\Bar{\boldsymbol{\delta}}_{i=\{7,8\}}$ \\
\hline
$xx$ & $\gamma_{1}$ & $-\frac{1}{2}(\gamma_{1}-3\gamma_2)$ & $\gamma_{1}$ & $\gamma_{1}$ & $-\frac{1}{2}(\gamma_{1}-3\gamma_2)$ & $\gamma_{1}$ & $\gamma_{3}$ \\
$yy$ & $\gamma_{2}$ & $\frac{1}{2}(3\gamma_{1}-\gamma_2)$ & $\gamma_{2}$ & $\gamma_{2}$ & $\frac{1}{2}(3\gamma_{1}-\gamma_2)$ & $\gamma_{2}$ & $\gamma_{3}$ \\
$zz$ & $\gamma_{4}$ & $\gamma_{4}$ & $\gamma_{4}$ & $\gamma_{4}$ & $\gamma_{4}$ & $\gamma_{4}$ & $\gamma_{5}$ \\
$xy$ & $\frac{\sqrt{3}}{2}(\gamma_1-\gamma_2)$ & $0$ & $-\frac{\sqrt{3}}{2}(\gamma_1-\gamma_2)$ & $\frac{\sqrt{3}}{2}(\gamma_1-\gamma_2)$ & $0$ & $-\frac{\sqrt{3}}{2}(\gamma_1-\gamma_2)$ & $0$ \\
$yx$ & $\frac{\sqrt{3}}{2}(\gamma_1-\gamma_2)$ & $0$ & $-\frac{\sqrt{3}}{2}(\gamma_1-\gamma_2)$ & $\frac{\sqrt{3}}{2}(\gamma_1-\gamma_2)$ & $0$ & $-\frac{\sqrt{3}}{2}(\gamma_1-\gamma_2)$ & $0$ \\
$xz$ & $\gamma_{6}$ & $0$ & $-\gamma_{6}$ & $\gamma_{7}$ & $0$ & $-\gamma_{7}$ & $0$ \\
$zx$ & $\gamma_{7}$ & $0$ & $-\gamma_{7}$ & $\gamma_{6}$ & $0$ & $-\gamma_{6}$ & $0$ \\
$yz$ & $\frac{\gamma_{6}}{\sqrt{3}}$ & $-\frac{2\gamma_{7}}{\sqrt{3}}$ & $\frac{\gamma_{6}}{\sqrt{3}}$ &
$\frac{\gamma_{7}}{\sqrt{3}}$ & $-\frac{2\gamma_{6}}{\sqrt{3}}$ & $\frac{\gamma_{7}}{\sqrt{3}}$ & $0$ \\
$zy$ & $\frac{\gamma_{7}}{\sqrt{3}}$ & $-\frac{2\gamma_{6}}{\sqrt{3}}$ & $\frac{\gamma_{7}}{\sqrt{3}}$ &
$\frac{\gamma_{6}}{\sqrt{3}}$ & $-\frac{2\gamma_{7}}{\sqrt{3}}$ & $\frac{\gamma_{6}}{\sqrt{3}}$ & $0$ 
\end{tabular}
\end{ruledtabular}
\end{table}

\begin{table}[ht]
\centering
\caption{Parameterization of $\Phi_{\mu\nu}(\Bar{\boldsymbol{R}}_{ji})/M$ for the next-nearest neighbors in positive $z$ direction.
}
\label{tab:phi_tau16}
\begin{ruledtabular}
\begin{tabular}{ccccccccccccc}
$\mu\nu\diagdown\Bar{\boldsymbol{R}}_{ji}$ &
$\Bar{\boldsymbol{\tau}}_{1}$ & $\Bar{\boldsymbol{\tau}}_{2}$ & $\Bar{\boldsymbol{\tau}}_{3}$ & $\Bar{\boldsymbol{\tau}}_{4}$ & $\Bar{\boldsymbol{\tau}}_{5}$ & $\Bar{\boldsymbol{\tau}}_{6}$  \\
\hline
$xx$ &
$\rho_{1}$ & $-\frac{1}{2}(\rho_2-3\rho_4)$ & $\rho_{1}$ & $\rho_{2}$ & $-\frac{1}{2}(\rho_1-3\rho_3)$ & $\rho_{2}$  \\

$yy$ &
$\rho_{3}$ & $\frac12(3\rho_2-\rho_4)$ & $\rho_{3}$ & $\rho_{4}$ & $\frac12(3\rho_1-\rho_3)$ & $\rho_{4}$ \\

$zz$ &
$\rho_{5}$ & $\rho_{6}$ & $\rho_{5}$ & $\rho_{6}$ & $\rho_{5}$ & $\rho_{6}$  \\

$xy$ &
$\frac{\sqrt{3}}{2}(\rho_1-\rho_3)$ & $0$ & $-\frac{\sqrt{3}}{2}(\rho_1-\rho_3)$ & $\frac{\sqrt{3}}{2}(\rho_2-\rho_4)$ & $0$ & $-\frac{\sqrt{3}}{2}(\rho_2-\rho_4)$  \\

$yx$ &
$\frac{\sqrt{3}}{2}(\rho_1-\rho_3)$ & $0$ & $-\frac{\sqrt{3}}{2}(\rho_1-\rho_3)$ & $\frac{\sqrt{3}}{2}(\rho_2-\rho_4)$ & $0$ & $-\frac{\sqrt{3}}{2}(\rho_2-\rho_4)$  \\

$xz$ &
$\rho_{7}$ & $0$ & $-\rho_{7}$ & $\rho_{10}$ & $0$ & $-\rho_{10}$  \\

$zx$ &
$\rho_{9}$ & $0$ & $-\rho_{9}$ & $\rho_{8}$ & $0$ & $-\rho_{8}$ \\

$yz$ &
$\frac{\rho_{7}}{\sqrt{3}}$ & $-\frac{2\rho_{10}}{\sqrt{3}}$ & $\frac{\rho_{7}}{\sqrt{3}}$ &
$\frac{\rho_{10}}{\sqrt{3}}$ & $-\frac{2\rho_{7}}{\sqrt{3}}$ & $\frac{\rho_{10}}{\sqrt{3}}$ \\

$zy$ &
$\frac{\rho_{9}}{\sqrt{3}}$ & $-\frac{2\rho_{8}}{\sqrt{3}}$ & $\frac{\rho_{9}}{\sqrt{3}}$ &
$\frac{\rho_{8}}{\sqrt{3}}$ & $-\frac{2\rho_{9}}{\sqrt{3}}$ & $\frac{\rho_{8}}{\sqrt{3}}$ 
\end{tabular}
\end{ruledtabular}
\end{table}

\begin{table}[ht]
\centering
\caption{Parameterization of $\Phi_{\mu\nu}(\Bar{\boldsymbol{R}}_{ji})/M$ for the next-nearest neighbors in negative $z$ direction.
}
\label{tab:phi_tau712}
\begin{ruledtabular}
\begin{tabular}{ccccccc}
$\mu\nu\diagdown\Bar{\boldsymbol{R}}_{ji}$ & 
$\Bar{\boldsymbol{\tau}}_{7}$ & $\Bar{\boldsymbol{\tau}}_{8}$ & $\Bar{\boldsymbol{\tau}}_{9}$ & $\Bar{\boldsymbol{\tau}}_{10}$ & $\Bar{\boldsymbol{\tau}}_{11}$ & $\Bar{\boldsymbol{\tau}}_{12}$ \\
\hline
$xx$ &
$\rho_{2}$ & $-\frac{1}{2}(\rho_1-3\rho_3)$ & $\rho_{2}$ & $\rho_{1}$ & $-\frac{1}{2}(\rho_2-3\rho_4)$ & $\rho_{1}$ \\

$yy$ &
$\rho_{4}$ & $\frac12(3\rho_1-\rho_3)$ & $\rho_{4}$ & $\rho_{3}$ & $\frac12(3\rho_2-\rho_4)$ & $\rho_{3}$ \\

$zz$ &
$\rho_{6}$ & $\rho_{5}$ & $\rho_{6}$ & $\rho_{5}$ & $\rho_{6}$ & $\rho_{5}$ \\

$xy$ &
$\frac{\sqrt{3}}{2}(\rho_2-\rho_4)$ & $0$ & $-\frac{\sqrt{3}}{2}(\rho_2-\rho_4)$ & $\frac{\sqrt{3}}{2}(\rho_1-\rho_3)$ & $0$ & $-\frac{\sqrt{3}}{2}(\rho_1-\rho_3)$ \\

$yx$ &
$\frac{\sqrt{3}}{2}(\rho_2-\rho_4)$ & $0$ & $-\frac{\sqrt{3}}{2}(\rho_2-\rho_4)$ & $\frac{\sqrt{3}}{2}(\rho_1-\rho_3)$ & $0$ & $-\frac{\sqrt{3}}{2}(\rho_1-\rho_3)$ \\

$xz$ &
$\rho_{8}$ & $0$ & $-\rho_{8}$ & $\rho_{9}$ & $0$ & $-\rho_{9}$ \\

$zx$ &
$\rho_{10}$ & $0$ & $-\rho_{10}$ & $\rho_{7}$ & $0$ & $-\rho_{7}$ \\

$yz$ &
$\frac{\rho_{8}}{\sqrt{3}}$ & $-\frac{2\rho_{9}}{\sqrt{3}}$ & $\frac{\rho_{8}}{\sqrt{3}}$ &
$\frac{\rho_{9}}{\sqrt{3}}$ & $-\frac{2\rho_{8}}{\sqrt{3}}$ & $\frac{\rho_{9}}{\sqrt{3}}$ \\

$zy$ &
$\frac{\rho_{10}}{\sqrt{3}}$ & $-\frac{2\rho_{7}}{\sqrt{3}}$ & $\frac{\rho_{10}}{\sqrt{3}}$ &
$\frac{\rho_{7}}{\sqrt{3}}$ & $-\frac{2\rho_{10}}{\sqrt{3}}$ & $\frac{\rho_{7}}{\sqrt{3}}$ 
\end{tabular}
\end{ruledtabular}
\end{table}

\section{Corrections}
Tables \ref{tab:phi_delta}, \ref{tab:phi_tau16}, and \ref{tab:phi_tau712} present corrected force constants as constrained by the symmetries of the P31m space group. By additionally choosing $\gamma_1=\gamma_2$, $\rho_1 = \rho_3$, $\rho_2=\rho_4$, $\rho_7 = \rho_9$, and $\rho_{8} = \rho_{10}$, Tables \ref{tab:phi_delta}, \ref{tab:phi_tau16}, and \ref{tab:phi_tau712} simplify to Table \ref{tab:force} in the original paper. For the updated results presented here, we continue to choose all next-nearest-neighbor force constant parameters equal; $\rho_i = \rho = -\gamma_1/10\sqrt{2}$. We also set $\gamma_2 = \gamma_1$ and $\gamma_7 = \gamma_6$. Note that these choices are not enforced by symmetry, but are made for simplicity.

\begin{figure}
    \centering
    \includegraphics[width=0.99\linewidth]{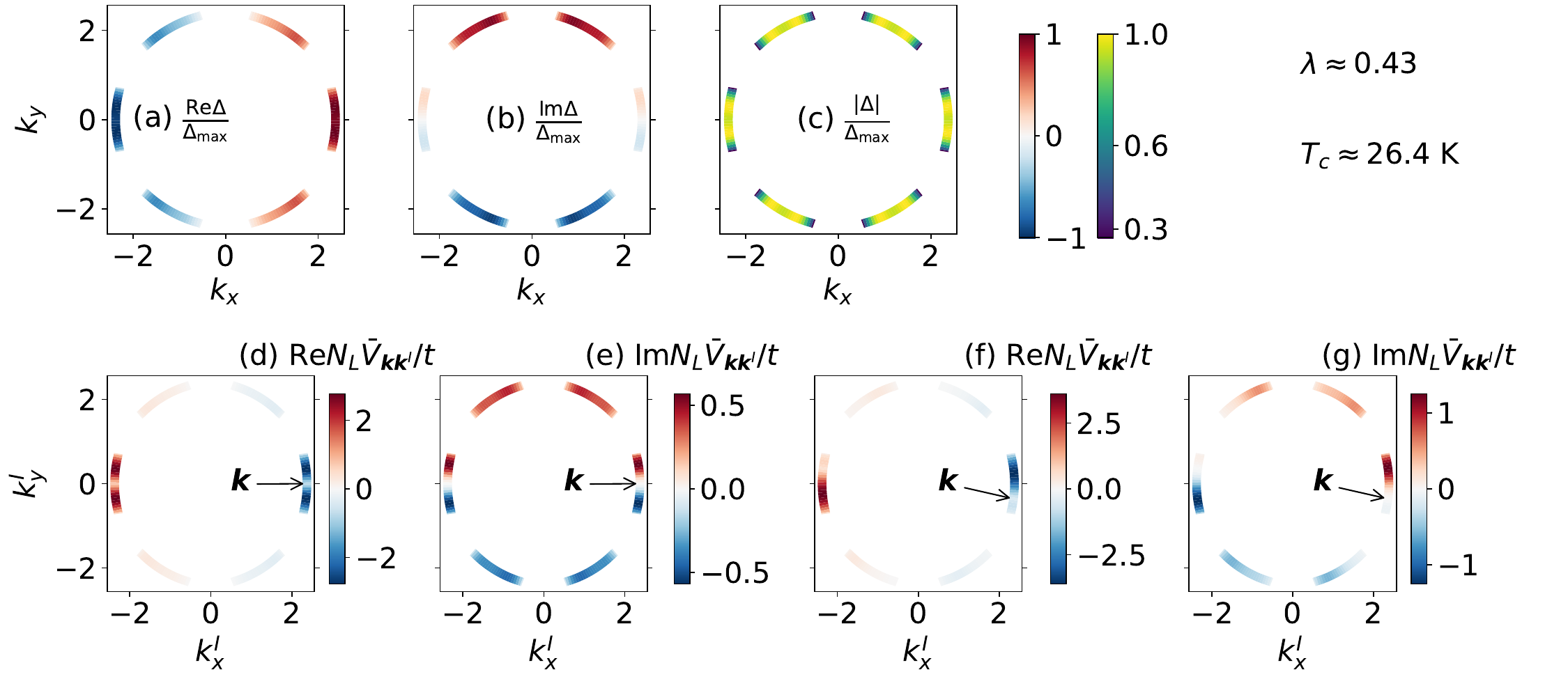}
    \caption{(a) Real, (b) imaginary, and (c) absolute value of the gap function on the bottom surface Fermi arc. The values are scaled by the maximum absolute value of the gap $\Delta_{\text{max}}$. The absolute value of the gap is the same on all six arcs. (d)-(g) give the phonon-mediated electron-electron interaction $\bar{V}_{\boldsymbol{k}\boldsymbol{k}'}$ as a function of $\boldsymbol{k}'$ with $\boldsymbol{k}$ fixed as indicated. The parameters are $t_o/t = 1.5$, $\beta/t = -1.5$,  $\mu/t = -0.05, \mu_o/t = 0.2, \alpha/t = -0.18$, $\gamma/t = -0.2$, $\gamma_1 = -(0.005t)^2$, $\gamma_3 = \gamma_5 = 0.45\gamma_1$, $\gamma_4 = \gamma_1$, $\gamma_6 = 1.5 \gamma_1$, $M = 4.87\times 10^{4}/t$, $\chi = 8$, $N_{\text{samp}} = 150$, and $L = 40$.}
    \label{fig:DVB}
\end{figure}

\begin{figure}
    \centering
    \includegraphics[width=0.8\linewidth]{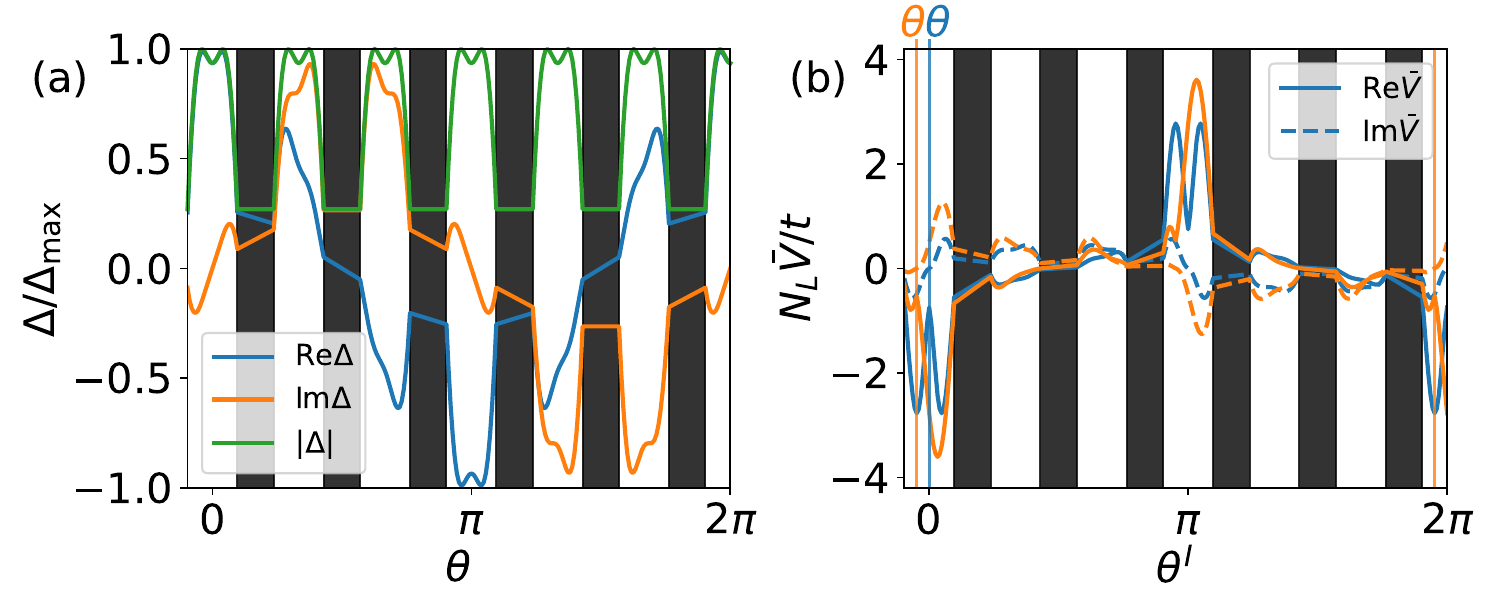}
    \caption{(a) The same as Fig.~\ref{fig:DVB}(a)-\ref{fig:DVB}(c), but now shown as a function of the angle $\theta$ that $\boldsymbol{k}$ makes with the $k_x$ axis. The black regions indicate directions where there is no Fermi arc. Here, we do not solve for the gap, and so the lines are just linearly interpolated. We show a bit more than one full revolution. (b) The phonon-mediated electron-electron interaction $\bar{V}_{\boldsymbol{k}\boldsymbol{k}'}$ as a function of the azimuthal angle $\theta'$ of $\boldsymbol{k}'$. The azimuthal angle $\theta$ for $\boldsymbol{k}$ is fixed as indicated for the blue and orange curves; in the center of a Fermi arc and the midpoint between the center and endpoint, respectively. The parameters are the same as Fig.~\ref{fig:DVB}.}
    \label{fig:DVBpt}
\end{figure}

\begin{figure}
    \centering
    \includegraphics[width=0.99\linewidth]{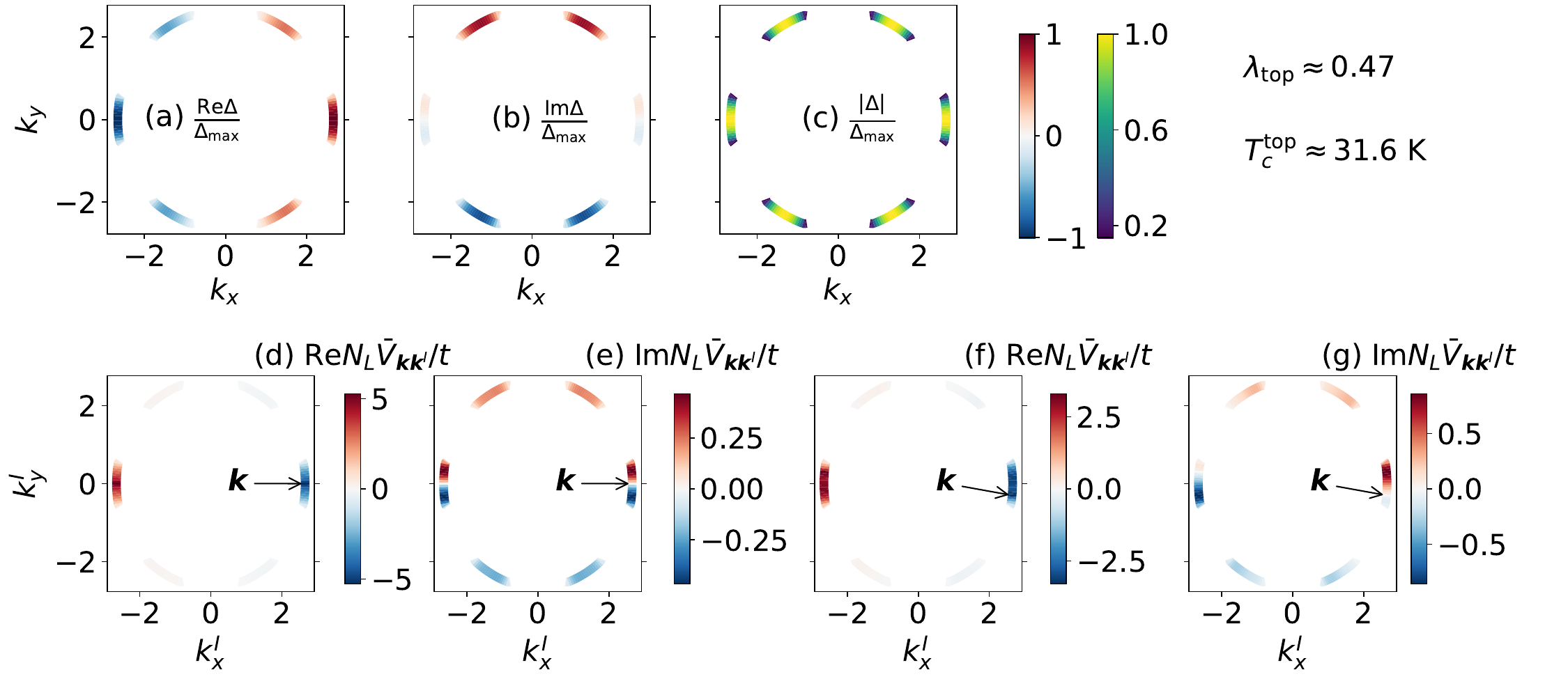}
    \caption{(a) Real, (b) imaginary, and (c) absolute value of the gap function on the top surface Fermi arc. The values are scaled by the maximum absolute value of the gap $\Delta_{\text{max}}$. The absolute value of the gap is the same on all six arcs. (d)-(g) give the electron-electron coupling $\bar{V}_{\boldsymbol{k}\boldsymbol{k}'}$ as a function of $\boldsymbol{k}'$ with $\boldsymbol{k}$ fixed as indicated. The parameters are $N_{\text{samp}} = 126$ and otherwise the same as Fig.~\ref{fig:DVB}.}
    \label{fig:DVT}
\end{figure}

\begin{figure}
    \centering
    \includegraphics[width=0.99\linewidth]{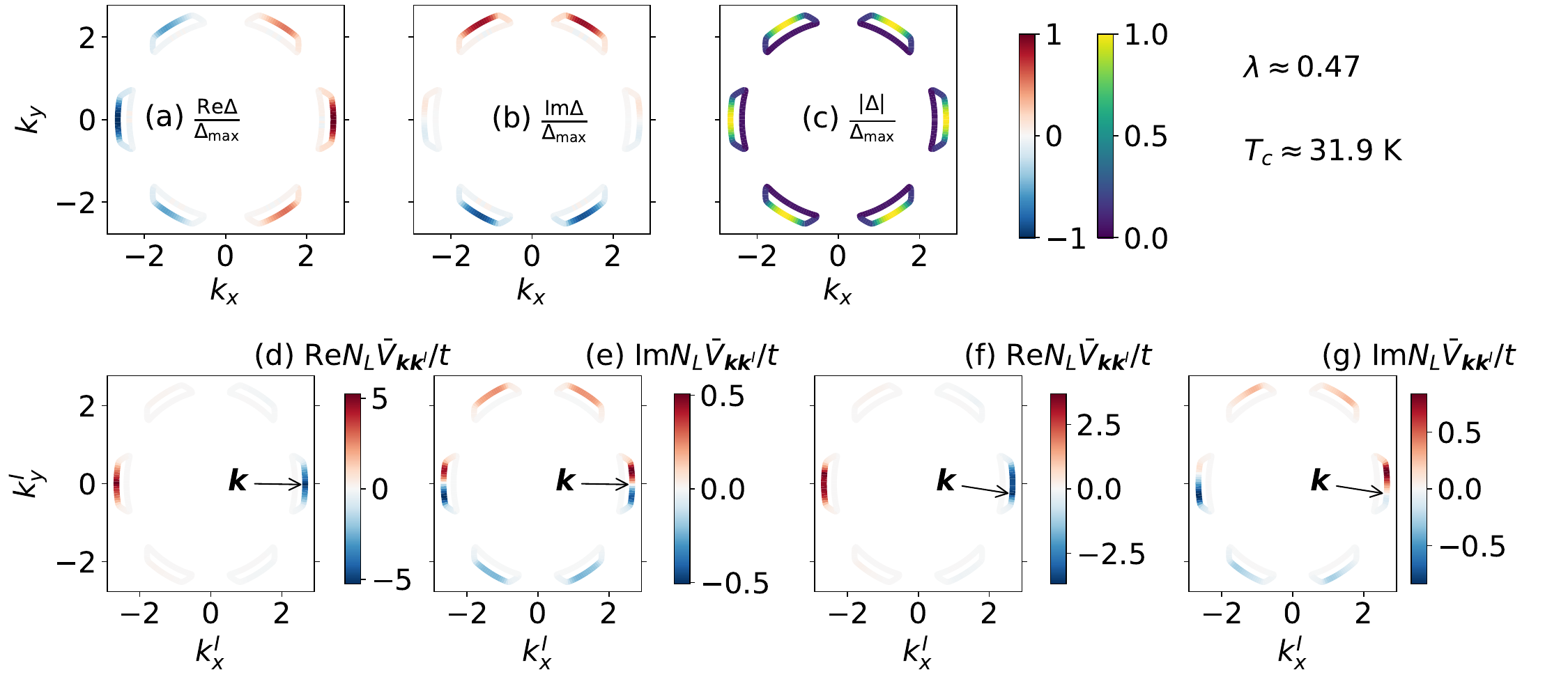}
    \caption{(a) Real, (b) imaginary, and (c) absolute value of the gap function on the full Fermi surface (FS). The values are scaled by the maximum absolute value of the gap $\Delta_{\text{max}}$. The absolute value of the gap is the same on all six pockets. (d)-(g) give the electron-electron coupling $\bar{V}_{\boldsymbol{k}\boldsymbol{k}'}$ as a function of $\boldsymbol{k}'$ with $\boldsymbol{k}$ fixed as indicated. The parameters are $N_{\text{samp}} = 330$ and otherwise the same as Fig.~\ref{fig:DVB}.}
    \label{fig:DVTBB}
\end{figure}

In the corrected results we use $Mt = 4.87\times 10^{4}$ where in the original paper we used $Mt = 5.6\times 10^{4}$. A lower mass slightly increases the strength of the electron-phonon coupling (EPC). $Mt = 4.87\times 10^{4}$ gives $M = 204$~u (the average atomic weight in PtBi$_2$) if $t=1$~eV and $a=1$~\AA. In the original paper we made the same claim about $Mt = 5.6\times 10^{4}$ due to a typo. We stress that Figs.~\ref{fig:OBCweight}, \ref{fig:phonon}, \ref{fig:gap}(f), \ref{fig:gap}(g), and \ref{fig:phononcompare} in the original paper remain correct as they do not rely on the electron-electron interaction.

Other than the mass, we present results with the same parameters as in the original paper in Figs.~\ref{fig:DVB}, \ref{fig:DVBpt}, \ref{fig:DVT}, and \ref{fig:DVTBB}.  The qualitative results are the same, but there are several quantitative changes. Figures \ref{fig:DVB} and \ref{fig:DVBpt} show the solution to the Fermi surface (FS) averaged linearized gap equation for the bottom surface Fermi arc. We still get a $p_x+ip_y$-wave gap and a local minimum of the gap amplitude in the center of the Fermi arc. The change is in the phonon-mediated electron-electron interaction. Unlike in the original paper, it now shows a local minimum of $\bar{V}_{\boldsymbol{k}\boldsymbol{k}'}$ at $\boldsymbol{k} = \boldsymbol{k}'$. That applies both for $\boldsymbol{k}$ in the center of the Fermi arc, and at the midpoint between the center and the endpoint. The local minimum in the gap amplitude in the center of the arc is due to strongest pairing when $\boldsymbol{k}$ and $\boldsymbol{k}'$ are at opposite sides of the center of the Fermi arc. That is most easily seen in Fig.~\ref{fig:DVBpt}(b).

In the original paper, we argue that $\bar{V}_{\boldsymbol{k}\boldsymbol{k}'}^{\text{FS}} = -\sum_m g_{\boldsymbol{k}\boldsymbol{k}'}^m g_{-\boldsymbol{k},-\boldsymbol{k}'}^m /\omega_{\boldsymbol{k}-\boldsymbol{k}',m} + \sum_m g_{\boldsymbol{k},-\boldsymbol{k}'}^m g_{-\boldsymbol{k},\boldsymbol{k}'}^m /\omega_{\boldsymbol{k}+\boldsymbol{k}',m}$ should have a peaked absolute value at $\boldsymbol{k}=\pm\boldsymbol{k}'$ due to the out-of-plane type EPC. For that contribution, $g_{\boldsymbol{k}\boldsymbol{k}}^m$ is nonzero and $\omega_{\boldsymbol{0},m}$ is small for the optical modes. That analytic expectation still stands, so the suppression of $\bar{V}_{\boldsymbol{k}\boldsymbol{k}'}$ at $\boldsymbol{k}=\pm\boldsymbol{k}'$ in Fig.~\ref{fig:DVBpt}(b) must be due to cancellations between all the terms that contribute to the sum.

That is confirmed by considering Fig.~\ref{fig:DVT} showing superconductivity on the top surface. There, $\bar{V}_{\boldsymbol{k}\boldsymbol{k}'}$ is peaked at $\boldsymbol{k}=\boldsymbol{k}'$, unlike the usual case for phonons. With the corrected calculation, it is now the top surface that dominates the superconductivity (has the largest $T_c$). We confirm that in Fig.~\ref{fig:DVTBB} with an updated version of Fig.~\ref{fig:gapFullFS} in the original paper. Now, we see that the gap is negligible outside the top surface Fermi arc on the full FS. For the top surface, we find a maximum gap amplitude in the center of the Fermi arc. That is because the phonon-mediated attraction is strongest in the center of the arc.

\begin{figure}
    \centering
    \includegraphics[width=0.99\linewidth]{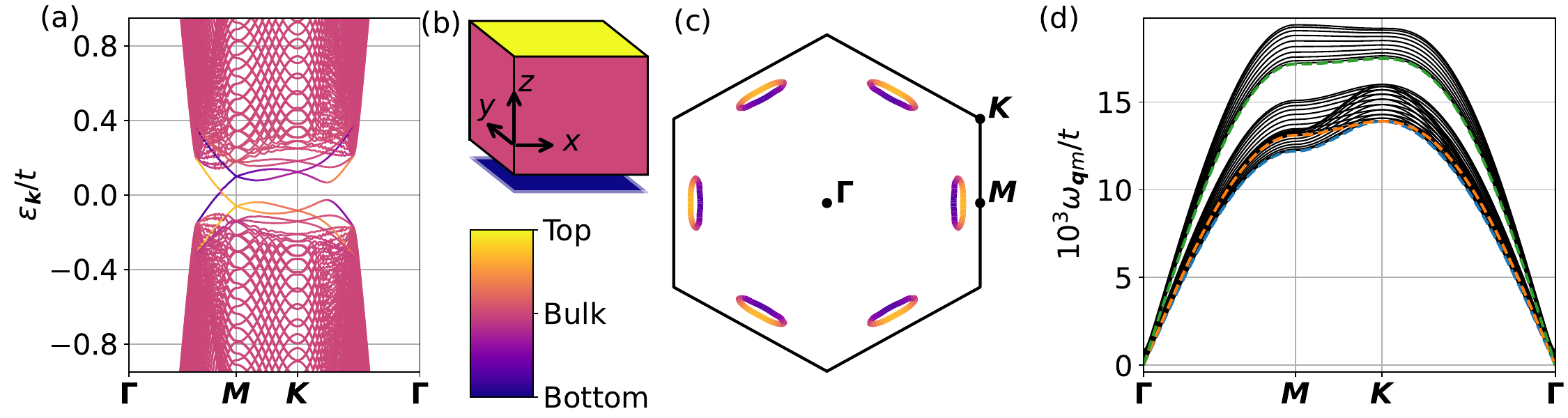}
    \caption{(a) Electron bands in a slab geometry, with each band colored by the weight of its eigenstate $W_{\boldsymbol{k}n}$. (b) Illustration of the system and the color code for top surface, bulk, and bottom surface. (c) FS shown within the first Brillouin zone, colored by the weight of the eigenstate. The full FS contains the Fermi arcs on the two surfaces and the projection of the bulk FS. (d) Black lines show the phonon spectrum in a slab geometry. Blue, orange and green dashed lines show the three acoustic phonon modes with periodic boundary conditions in all directions at $q_z = 0$. The parameters are $t_o/t = 0.05$, $\beta/t = -1.5$, $\beta_o/t = -0.1$, $\mu/t = -0.02, \mu_o/t = -1.3, \alpha/t = -0.08$, $\gamma/t = -0.08$, $\gamma_1 = -(0.005t)^2$, $\gamma_3 = 0.43\gamma_1$, $\gamma_4 = 1.5\gamma_1$, $\gamma_5 = 0.432\gamma_1$, $\gamma_6 = 0.6 \gamma_1$, (a),(c) $L = 45$, and (d) $L=10$. }
    \label{fig:banderr}
\end{figure}

We obtain that $\bar{V}_{\boldsymbol{k}\boldsymbol{k}'}$ is peaked at $\boldsymbol{k}=\boldsymbol{k}'$ on both surfaces by tuning the phonon parameters, specifically we reduce the phonon energy at $\boldsymbol{q} =0$ and increase it at $\boldsymbol{q}=\boldsymbol{M}$ compared to the original paper. See Fig.~\ref{fig:banderr}(d) for the phonon spectrum at new parameters. In that case, we find that the most common behavior is to have maximum gap amplitude in the center of the Fermi arc on both surfaces.

\begin{figure}
    \centering
    \includegraphics[width=0.99\linewidth]{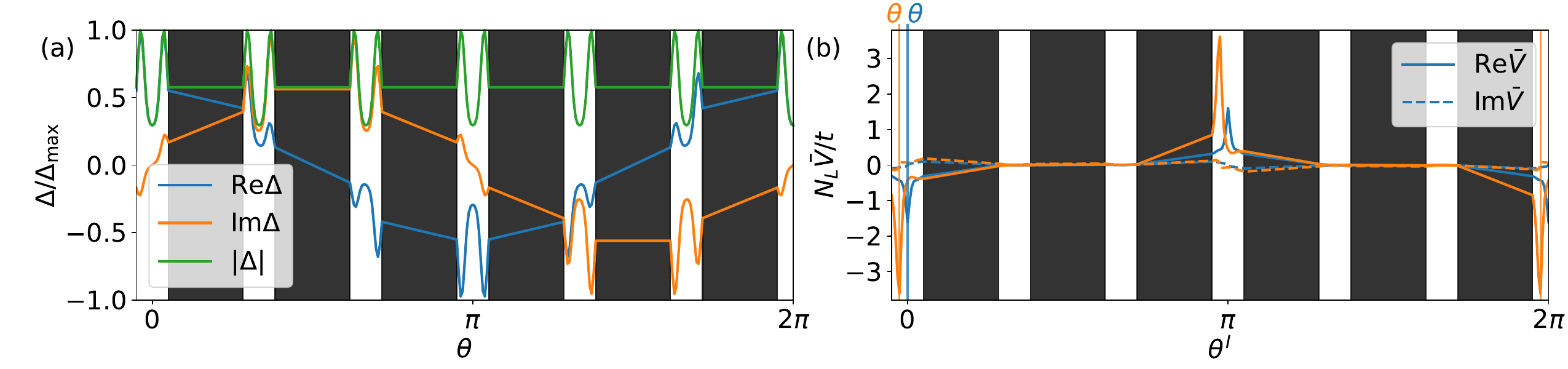}
    \caption{(a) Real, imaginary, and absolute value of the gap function on the bottom surface Fermi arc, which dominates the superconductivity at the new parameters. The gaps are shown as a function of the angle $\theta$ that $\boldsymbol{k}$ makes with the $k_x$ axis, still giving us a $p_x+ip_y$-wave gap function in the band basis. The black regions indicate directions where there is no Fermi arc. (b) The phonon-mediated electron-electron interaction $\bar{V}_{\boldsymbol{k}\boldsymbol{k}'}$ as a function of the azimuthal angle $\theta'$ of $\boldsymbol{k}'$. The azimuthal angle $\theta$ for $\boldsymbol{k}$ is fixed as indicated for the blue and orange curves; in the center of a Fermi arc and the midpoint between the center and endpoint, respectively.  The parameters are $M = 4.87\times 10^{4}/t$, $\chi = 14$, $N_{\text{samp}} = 102$, $L = 45$, and otherwise the same as Fig.~\ref{fig:banderr}. We find $\lambda \approx 0.34$ giving $T_c \approx 13.7$~K if $t = 1$~eV.}
    \label{fig:DVBptErr}
\end{figure}

A natural question is if it is still possible to find a local minimum of the gap in the center of the Fermi arc as presented in the original paper when $\bar{V}_{\boldsymbol{k}\boldsymbol{k}'}$ is peaked at $\boldsymbol{k}=\boldsymbol{k}'$ on both surfaces. For that, we find it necessary to introduce a separate hopping strength for intra- and interorbital hopping out of plane. We label these $\beta$ and $\beta_o$, respectively. Then, the upper diagonal block in the slab Hamiltonian matrix is
\begin{equation}
    H_U(\boldsymbol{k}) = \frac{1}{2}\begin{pmatrix}
         \beta & 0 & \beta_o & -s_{\boldsymbol{k}}  \\
         0 &\beta & -s_{\boldsymbol{k}}^* & \beta_o  \\
         -\beta_o & -s_{\boldsymbol{k}} & -\beta & 0  \\
         -s_{\boldsymbol{k}}^* & -\beta_o & 0 & -\beta 
    \end{pmatrix}.
\end{equation}
With the hierarchy of hopping amplitudes $|t_o| < |\beta_o| < t  < \beta$, we find a local minimum of the gap in the center of the Fermi arc in Fig.~\ref{fig:DVBptErr}(a). From Fig.~\ref{fig:DVBptErr}(b) we see that the reason is that here, the phonon-mediated attraction $\bar{V}_{\boldsymbol{k}\boldsymbol{k}}$ is weaker in the center of the arc than at the midpoint between the center and the endpoint of the arc. The new electron bands and full FS are shown in Figs.~\ref{fig:banderr}(a) and \ref{fig:banderr}(c). Note that compared to the parameters used in the original paper, the surface states are less extremely surface localized, compare Fig.~\ref{fig:OBCweight}(c) in the original paper and Fig.~\ref{fig:banderr}(c). Also, the two surface states that cross between $\boldsymbol{\Gamma}$ and $\boldsymbol{M}$ in Fig.~\ref{fig:banderr}(a) seem to show some degree of hybridization. In other words, there are finite size effects, unlike the case at the parameters used in the original paper. 

Remember that the weight of the states is defined as $W_{\boldsymbol{k}n} = \sum_{z_i = 1}^{L} (z_i-1)|\psi_{\boldsymbol{k}n,z_i}|^2/(L-1),$
where $|\psi_{\boldsymbol{k}n,z_i}|^2$ is the sum of squares of the four entries in the eigenvector of band $n$ associated with layer $z_i$. By considering $|\psi_{\boldsymbol{k}n,z_i}|^2$ for the bands with Fermi arcs and with momentum in the center of the Fermi arc, we find that as a function of $z_i$, $|\psi_{\boldsymbol{k}n,z_i}|^2$ shows a smooth exponential decay from the surface at the parameters in Fig.~\ref{fig:DVB}. The occupation on the layer next to the surface is not negligible, and so the out-of-plane contribution to EPC is not suppressed in the center of the Fermi arc. At the parameters used in Fig.~\ref{fig:banderr}, on the other hand, $|\psi_{\boldsymbol{k}n,z_i}|^2$ shows a jagged behavior. The electrons are spread on more layers close to the surface, but at the same time certain layers have a very small, almost negligible occupation compared to their neighbors. We conjecture that this suppresses the contribution from the out-of-plane type EPC in the center of the arc. As the momentum moves along the arc, the occupation on the previously suppressed layers increases, and so does the strength of EPC. Eventually, when approaching the end of the Fermi arc the states become more bulk-like, and then the strength of the phonon-mediated electron-electron attraction decreases again. Jagged occupation on layers could be the case in PtBi$_2$ due to its trilayer structure \cite{Shipunov2020ExpWeylSC, Vocaturo2024PtBi2Effective}. A jagged occupation on layers can also occur in other Weyl semimetals, an example being the diamond lattice model in Ref.~\cite{Okugawa2014FermiArcEffective}. 

Additionally, we note that the hopping hierarchy $|t_o| < |\beta_o| < t  < \beta$ could be reasonable for PtBi$_2$. The orbitals in the effective model are unspecified, but the signs of hoppings indicate that one has even and the other has odd parity. In PtBi$_2$, $p$- and $d$-wave orbitals dominate \cite{Vocaturo2024PtBi2Effective}. Thus, it seems reasonable that interorbital hopping is smaller than intraorbital hopping. In general, for multiorbital systems, $\beta_o = \beta$ is less likely than $\beta_o \neq \beta$, so this change of the model is more realistic.

With calculations such as Ref.~\cite{Maeland2026Jul}, we briefly consider the role of Coulomb repulsion for the case of large surface state bandwidth compared to the maximum phonon energy. If EPC prefers a local minimum of the gap at the center of the Fermi arcs, adding Coulomb repulsion decreases the gap amplitude there relative to the gap maximum. That is because the Coulomb repulsion is strongest in the center of the Fermi arcs.

Finally, Eqs.~\eqref{eq:Vtaylor} and \eqref{eq:Vphi} in the original paper should read
\begin{align}
    V = \sum_{i\alpha \neq j \beta} V(\Bar{\boldsymbol{r}}_{j\beta}-\Bar{\boldsymbol{r}}_{i\alpha}) &\approx \sum_{i\alpha \neq j \beta} \bigg[ V(\Bar{\boldsymbol{R}}_{j\beta}-\Bar{\boldsymbol{R}}_{i\alpha}) + \sum_\mu \left. \pdv{V}{r_{i\alpha\mu}}\right\rvert_{\text{eq}} u_{i\alpha\mu} +\frac{1}{2} \sum_{\mu\nu} \left.\frac{\partial^2 V}{\partial r_{i\alpha\mu} \partial r_{j\beta\nu}}\right\rvert_{\text{eq}}u_{i\alpha\mu}  u_{j\beta\nu} \bigg],
\end{align}
\begin{align}
    V &= \frac{1}{2}\sum_{i\alpha\mu, j \beta\nu} \left.\frac{\partial^2 V}{\partial r_{i\alpha\mu} \partial r_{j\beta\nu}}\right\rvert_{\text{eq}}u_{i\alpha\mu}  u_{j\beta\nu} = \frac{1}{2}\sum_{i\alpha\mu, j \beta\nu} \Phi_{\mu\nu}^{\alpha\beta}(\Bar{\boldsymbol{R}}_j-\Bar{\boldsymbol{R}}_i) u_{i\alpha\mu}  u_{j\beta\nu},
\end{align}
i.e., while the derivatives are evaluated in equilibrium, the derivatives of $V(\Bar{\boldsymbol{r}}_{j\beta}-\Bar{\boldsymbol{r}}_{i\alpha})$ should be taken with respect to the instantaneous positions $\Bar{\boldsymbol{r}}_{i\alpha}$ and $\Bar{\boldsymbol{r}}_{j\beta}$. These typos had no consequence as we did not model the potential energy $V(\Bar{\boldsymbol{r}}_{j\beta}-\Bar{\boldsymbol{r}}_{i\alpha})$ nor compute these derivatives to derive the phonon spectrum.

\section{Acknowledgments}
We thank Kristoffer Leraand for pointing out the error in our original calculation.

\end{document}